\documentclass[manuscript]{aastex}
\usepackage{tikz}
\usetikzlibrary{snakes}

\slugcomment{Not to appear in Nonlearned J., 45.}

\shorttitle{The Tale of Two Emergences}
\shortauthors{Centeno et al.}

\begin{document}


\title{A Tale of Two Emergences: Sunrise II Observations of Emergence Sites in a Solar Active Region}





\author{\textsc{
R.~Centeno,$^{1}$
J.~Blanco~Rodr\'{\i}guez,$^{2}$
J.~C.~Del~Toro~Iniesta,$^{3}$
S.~K.~Solanki,$^{4,5}$
P.~Barthol,$^{4}$
A.~Gandorfer,$^{4}$
L.~Gizon,$^{4, 6}$
J.~Hirzberger,$^{4}$
T.~L.~Riethm\"uller,$^{4}$
M.~van~Noort,$^{4}$
D.~Orozco~Su\'arez,$^{3}$
T.~Berkefeld,$^{7}$
W.~Schmidt$^{7}$
V.~Mart\'{\i}nez Pillet,$^{8}$
\& M.~Kn\"olker,$^{1}$
}}

\affil{
$^{1}$High Altitude Observatory, National Center for Atmospheric Research, P.O. Box 3000, Boulder, CO 80307-3000, USA\\
$^{2}$Grupo de Astronom\'{\i}a y Ciencias del Espacio, Universidad de Valencia, 46980 Paterna, Valencia, Spain\\
$^{3}$Instituto de Astrof\'{\i}sica de Andaluc\'{\i}a (CSIC), Apartado de Correos 3004, 18080 Granada, Spain\\
$^{4}$Max-Planck-Institut f\"ur Sonnensystemforschung, Justus-von-Liebig-Weg 3, 37077 G\"ottingen, Germany\\
$^{5}$School of Space Research, Kyung Hee University, Yongin, Gyeonggi, 446-701, Republic of Korea\\
$^{6}$Institut f\"ur Astrophysik, Georg-August-Universit\"at G\"ottingen, Friedrich-Hund-Platz 1, 37077 G\"ottingen, Germany\\
$^{7}$Kiepenheuer-Institut f\"ur Sonnenphysik, Sch\"oneckstr. 6, 79104 Freiburg, Germany\\
$^{8}$National Solar Observatory, 3665 Discovery Drive, Boulder, CO 80303, USA\\
}

\begin{abstract}
In June 2013, the two scientific instruments onboard the second {\sc Sunrise} mission witnessed, in detail, a small-scale magnetic flux emergence event as part of the birth of an active region.
The Imaging Magnetograph Experiment (IMaX) recorded two small ($\sim 5''$) emerging flux patches in the polarized filtergrams of a photospheric Fe {\sc i} spectral line. Meanwhile, the Sunrise Filter Imager (SuFI) captured the highly dynamic chromospheric response to the magnetic fields pushing their way through the lower solar atmosphere. The serendipitous capture of this event offers a closer look at the inner workings of active region emergence sites. In particular, it reveals in meticulous detail how the rising magnetic fields interact with the granulation as they push through the Sun's surface, dragging photospheric plasma in their upward travel. The plasma that is burdening the rising field slides along the field lines, creating fast downflowing channels at the footpoints.
The weight of this material anchors this field to the surface at semi-regular spatial intervals, shaping it in an undulatory fashion. Finally, magnetic reconnection enables the field to release itself from its photospheric anchors, allowing it to continue its voyage up to higher layers. This process releases energy that lights up the arch-filament systems and heats the surrounding chromosphere.
\end{abstract}


\section{Introduction}

Magnetic fields on the Sun operate at a large range of spatial and temporal scales, and all of them play a fundamental role in its structure, its dynamics, and its energy budget. 
These fields connect the inside of the Sun with its atmosphere providing means for storing, transporting, and depositing energy among the different atmospheric layers. They also draw pathways throughout the interplanetary medium, magnetically linking the Sun to the Earth and the entire solar system. Magnetic fields are at the core of the solar cycle and they are the seeds of space weather.

Active regions (ARs) are the most visible manifestation of magnetic fields on the Sun. They appear on the solar surface in the form of magnetic dipoles. It is generally believed that the magnetic field of ARs emerges in the form of $\Omega$-loops whose rise is triggered by deep convective flows and buoyant instabilities in a toroidal field located near the base of the convection zone \citep{parker1955}. 
Although this paradigm explains many of the observed aspects of ARs and the solar cycle, these flux tubes do not emerge as a whole; rather, they break through the photosphere in small bundles \citep[see, e.g.][]{birch}, slowly piling up magnetic flux at the surface, that then aggregates into stronger concentrations and gives rise to sunspots \citep{cheung2010}. 
\cite{rolf} describe observational evidence of small-scale emergence of magnetic flux (associated to elongated granulation) as the main contributor to the overall flux of an AR.
This is the picture of {\em resistive emergence} described by \cite{pariat2004}, in which the $\Omega$-loops rising from the convection zone develop spatial undulations, whose crests emerge because of the Parker instability, while their troughs remain trapped below the surface, until they are released by magnetic reconnection. In their work, these locations are often associated with Ellerman Bombs \citep[EB,][]{ellerman} and bright points. 
The serpentine nature of the emerging fields results in small-scale moving dipolar features \citep[MDFs,][]{bernasconi2002} of opposite polarity to that of the AR. These are the footpoints of U-loops where the serpentine field is anchored below the surface due to trapped mass, and become the channels where the material that was carried up by the crests of the undulating field, flows back down into the surface \citep[see][]{cheung2010, centeno2012}.  

When studying the small-scale details in AR emergence sites, \cite{ortiz14} reported semi-circular magnetic footpoints, straddling several granules, braketing patches of horizontal field in the photosphere. Chromospheric spectral analysis showed that these footpoints always surround a bubble of cool chromospheric material that increases in size as it ascends into higher layers. Their results are in qualitative agreement with numerical simulations of flux emergence \citep{juan2008, juan2009}.
In a subsequent paper, \cite{jaime2015} show that these granular sized magnetic bubbles emerge up to the chromosphere, where the pre-exisiting field seems to hinder their further ascent. Signatures of heating around the edges of the bubble are reported, but no clear physical mechanisms are identified.
Recent joint observations with the Swedish Solar Telescope, the Interface Region Imaging Spectrograph (IRIS), and Hinode have revealed the transition region response to these small-scale events in the emergence sites of ARs. Upward velocities predominate in the emergence sites and, when the orientation of the emerging flux differs from that of the overarching filament system, brightenings ensue. The transition region response is typically delayed by 10 minutes with respect to the photospheric emergence \citep{ortiz16}.

Magnetic reconnection is one of the most likely mechanisms responsible for heating the upper atmospheric layers. The reconnection process leads to energy conversion, from magnetic into thermal and kinetic energy \citep{priest2000}, and it  shows up in the observations in the form of brightenings and/or surges \citep{yokoyama}. Ellerman bombs occur in areas where intricate magnetic topologies are likely to lead to magnetic reconnection \citep{georgoulis2002}, such as places where newly emerged magnetic flux interacts with preexisting ambient fields, the locations above MDFs and the neutral lines surrounding the boundaries of supergranular cells.
Although there is some controversy in the literature of what constitutes an EB \citep{vissers2015, rutten2016}, they can be grouped in a category of events that involve magnetic reconnection and subsequent heating of mid-chromospheric layers and tend to happen rather frequently in the emergence site of ARs.

 In this work we study, at a constant, high resolution, the detailed structure and evolution of the emergence site of a developing active region using data from the second flight of the {\sc Sunrise} mission. The scientific payload onboard {\sc Sunrise II} captured two small flux emergence events in meticulous detail, registering the photospheric and chromospheric responses to the incipient magnetic field that rises from below the surface to become part of the larger-scale active region.
The instruments and the observations used in this work are presented in Section \ref{observations}. In Section \ref{photosphere}, we delve into a detailed description of the photospheric signatures of the emerging magnetic flux, whilst Section \ref{chromosphere} analyzes the chromospheric response to these events. All of the information gathered from the observations is then folded into a coherent picture discussed in Section \ref{discussion}, followed by speculations and final remarks.

\section{Observations}\label{observations}

\begin{figure}[!t]
\includegraphics[angle=0,scale=.55]{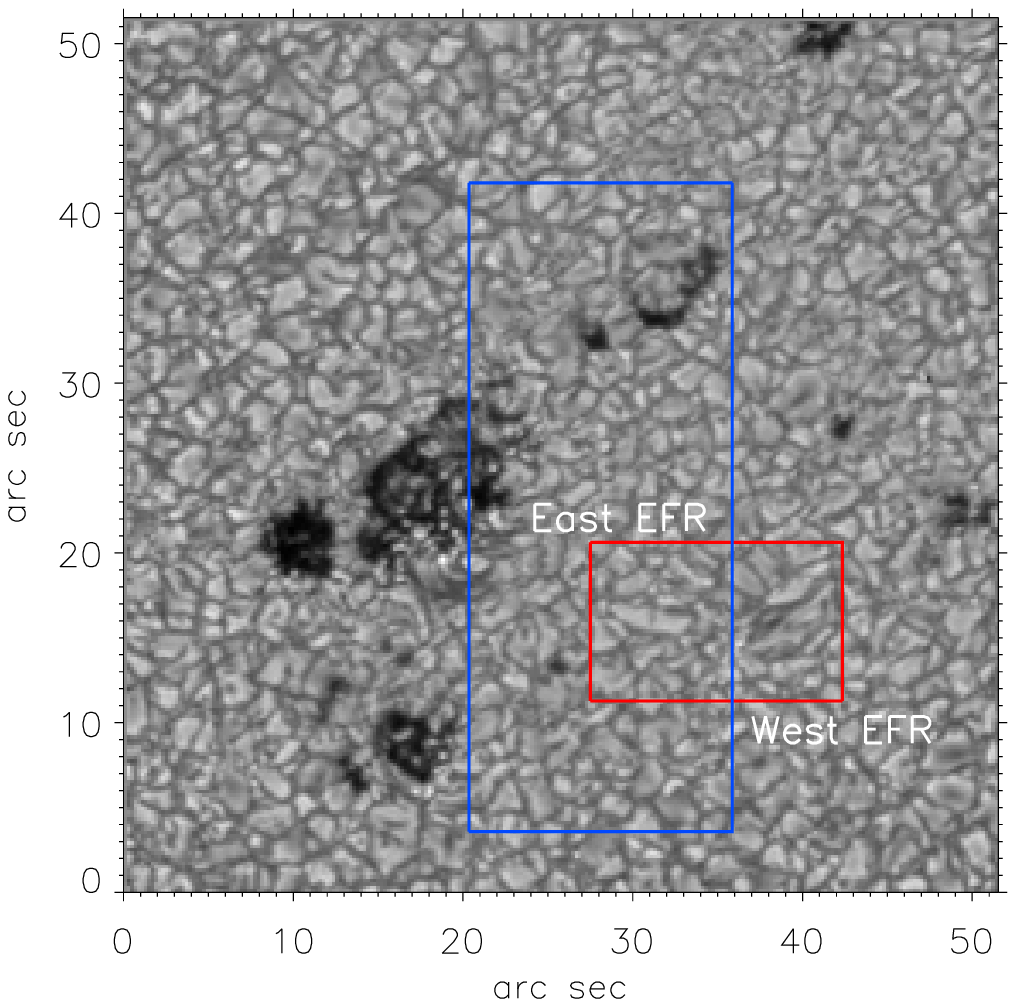}
\includegraphics[angle=0,scale=0.55]{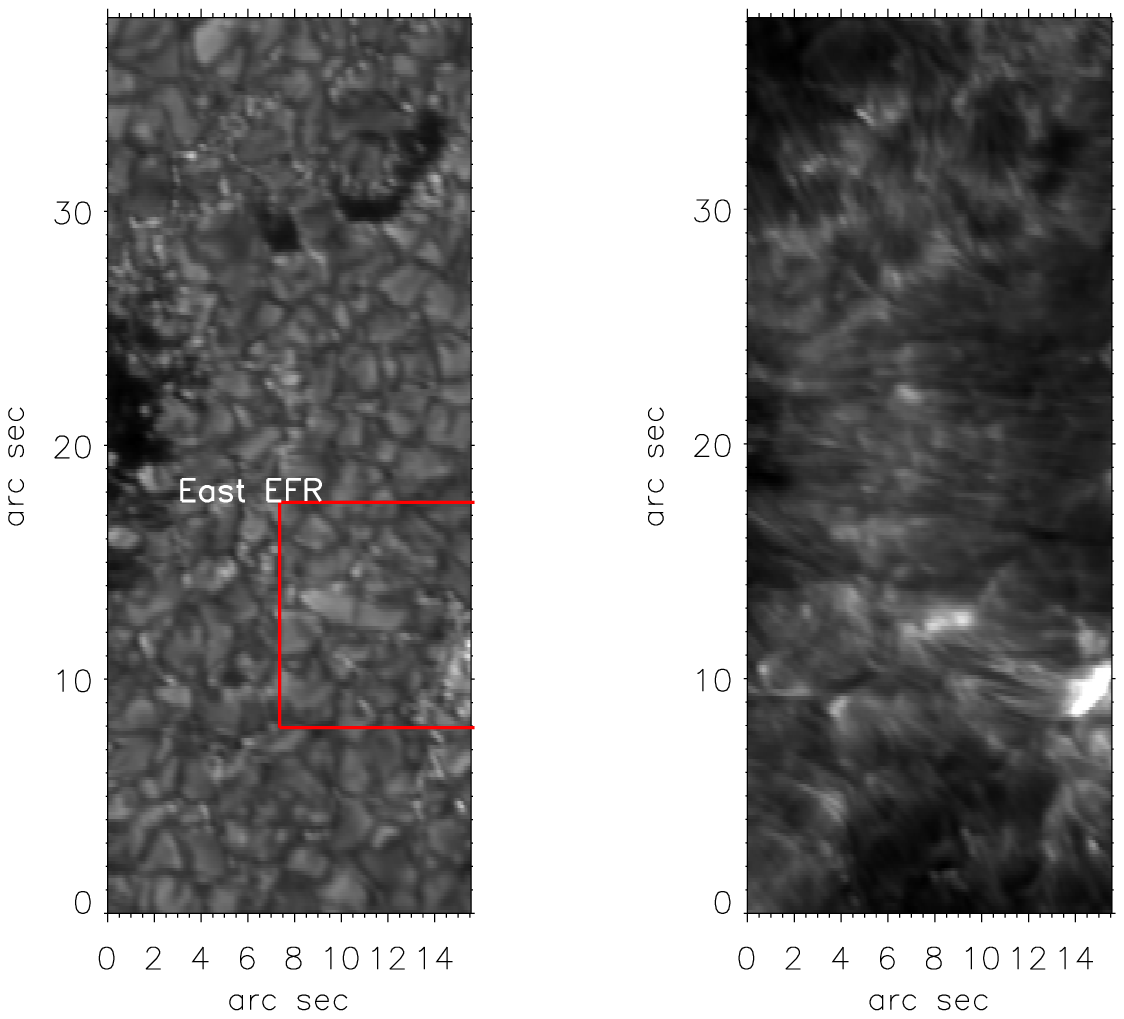}
\caption{Left: Reconstructed continuum image from the IMaX instrument. The blue rectangle represents the smaller FOV of the SUFI instrument, whilst the red rectangle delimits the area of interest for this work. Within it, two small flux emergence events (labeled `East EFR' and `West EFR', respectively) show the typical signatures of elongated granulation. Middle and right panels: images from SUFI in 2 wavelength bands, at 300 (UV continuum) and 397 nm (Ca {\sc ii} H). Notice how the FOV of SuFI only captures the East EFR and the confluence point between both magnetic patches.\label{fig:context}}
\end{figure}

The data presented in this paper were obtained during the first day of the second flight of the balloon-borne  mission {\sc Sunrise II} \citep{solanki2016}. {\sc Sunrise II} was launched from Kiruna (Sweden) on 2013 June 12, and it flew for 5 consecutive days at an average flight altitude of 35~km. This granted the possibility of a 24~hr duty cycle in a near-space environment, which minimized turbulence perturbations and allowed for ultraviolet (UV) observations. The scientific payload was similar to that of the first flight \citep{bar11, berkefeld2011}, comprising the Sunrise Ultraviolet Filter Imager \citep[SuFI,][]{gan11} and the Imaging Magnetograph Experiment \citep[IMaX,][]{pillet11}.
Two main data reduction levels were produced for each of the instruments: Level 1.x refers to calibrated data prior to deconvolution with the instrument PSF, whereas Level 2.x data were subjected to a spatial reconstruction. Despite the higher contrast of the latter, the increased noise and the small artifacts that arise from the deconvolution process render Level 2.x data risky for the analysis of weak polarization signals. Throughout this paper we make use of Level 1.2 data (see below) in all quantitative aspects of the analysis, relegating the reconstructed data to visualization purposes only. This comes at the expense of spatial resolution.

After refurbishment, the IMaX instrument on {\sc Sunrise II} remained essentially unchanged with respect to its first flight \citep{pillet11, solanki2016}. IMaX uses a LiNbO$_3$ solid etalon for spectral analysis and liquid crystal variable retarders for polarization modulation. 
The data presented in this paper were acquired using the so-called \emph{V8-4} observing mode, which produces the full Stokes vector ($I$, $Q$, $U$, $V$) at 8 wavelengths around the magnetically sensitive Fe {\sc i} 525.0208 nm line, which has a Land\'e factor of 3 and forms in the photosphere of the Sun.
 The spectral region was sampled at $[-12, -8, -4, 0, 4, 8, 12, 22.7]$ pm with respect to the reference wavelength, yielding 7 wavelength samples within the line and one in the red continuum. In order to improve the signal to noise ratio, 4 accumulations were taken for each polarization modulation state and each wavelength position. A whole scan operation in this observing mode takes approximately 36.5~s. The time series used in this paper is composed of 28 individual observation cycles and is $\sim 17$ minutes long.
The total field of view (FOV) covers $51^{\prime\prime} \times 51^{\prime\prime}$ with a pixel size of $0.055^{\prime\prime}$. 

\noindent A further set of treatments was applied to the level 1 data used for this study. First, the whole time series was carefully aligned to correct for remaining jitter. Second, a temporal interpolation was applied to the images in order to compensate for the non-simultaneity of the measurements in each cycle. As noted above, an IMaX observation cycle takes approximately 36.5 s, during which the instrument tunes through the different wavelength positions and polarization modulation states. The aim of this interpolation is to put the whole array of data into a common time frame, thus reducing the artifacts introduced by solar evolution. 
Lastly, a wavelength correction across the FOV was performed to compensate for the wavelength blueshift introduced by the IMaX etalon in its collimated configuration \citep[see, e.g.,][]{puschmann06,reardon08,pillet11}. In order to correct for this effect, the data reduction pipeline includes a model of the wavelength blueshift that is obtained from a surface fit to a third-degree polynomial over the FOV of a flat-field image. 
All of these treatments led to the production of the Level 1.2 data used in this paper.

Simultaneous measurements from the SuFI instrument were carried out in a subset of the IMaX field of view (delimited by the blue rectangle in the left panel of Fig. \ref{fig:context}) at a high cadence. SuFI cycled through 3 broadband wavelength channels (i.e. a 5\,nm wide channel in the UV continuum around 300 nm and two narrower bands of 0.11 nm and 0.18 nm around the the Ca {\sc ii} H line at 396.8 nm) in approximately 7~s \citep{riethmueller2013}, producing brightness images of the photosphere and the low chromosphere in a portion of the IMaX FOV, with a 
$\sim 0.02^{\prime\prime}$/px spatial sampling. A sample of the images of a SuFI cycle is shown in the middle and right panels of Figure \ref{fig:context}. 

The data used for this work were obtained on 2013 June 12, starting at 23:39 UT and spanning $\sim$17 minutes. The telescope pointed to the newly emerged active region NOAA AR 11768, which was located at  $\sim 11^{\circ}$ S, $\sim 16^{\circ}$ W in solar coordinates. The larger FOV of the IMaX instrument (left panel of Fig. \ref{fig:context}) covered the emergence site and the trailing polarity of the AR, which comprised several pores and a penumbra-like structure.
This paper focuses on a particular section of the FOV, delimited by the red rectangle in Figure~\ref{fig:context}. During the time series, this area shows two separate instances of magnetic field emergence (evidenced by the  elongated granulation in the continuum image) that contribute to the overall emergence of magnetic field in the AR. At the time of the snapshot in this figure, the two emerging flux regions (EFR) are already present and clearly visible. Throughout this paper we refer to them as {\em East EFR} and {\em West EFR}, whilst the location in between them, where the opposite polarity footpoints of the two EFRs meet, is referred to as the {\em confluence point} (enlarged images of the FOV of the red rectangle can be seen in Figure \ref{fig:evolution}). In this paper, we carry out a detailed study of the temporal evolution of the magnetic and thermodynamic properties in the photosphere as well as the chromospheric response to the newly emerged flux.
 
Data from the Helioseismic and Magnetic Imager \citep[HMI,][]{hmi} on board the Solar Dynamics Observatory \citep[SDO,][]{sdo} were used as context to understand the evolution of the AR before and after the {\sc Sunrise II} observations. Data from the Atmospheric Imaging Assembly \cite[AIA,][]{aia_paper} were also used to study other facets of the chromospheric response to the emerging flux.

\subsection{Spectral line inversion and p-mode filtering of IMaX data} \label{inversions}

In order to obtain the magnetic properties of the emerging flux site at the photosphere we proceeded to carry out the spectral line inversion of the spectropolarimetric data from the IMaX instrument. For this purpose we used a  Milne-Eddington (ME) spectral line inversion code\footnote{See reference to HEXIC in http://www2.hao.ucar.edu/csac/csac-spectral-line-inversions.}, generalized from the Very Fast Inversion of the Stokes Vector \citep[VFISV,][]{vfisv, vfisv2}. The ME approximation solves the polarized radiative transfer equation assuming that the thermodynamical and magnetic properties of the atmosphere are height independent, except for the source function, which varies linearly with optical depth. The generation of spectral line polarization is taken to be exclusively due to the Zeeman effect. The inversion algorithm itself uses the Levenberg-Marquardt technique \citep{press} for non-linear least-squares minimization of a merit function that quantifies the difference between the synthetic and the observed spectral lines. All inversions were performed by fixing the magnetic filling factor to 1, but they were ran twice, using two different values for the stray light fraction, $0\%$ and $25\%$ (the latter being the approach taken by the IMaX data calibration team). This was done in order to explore the quantitative and qualitative differences in the results between these two extremes. We found that the transverse component of the magnetic field increased by $\sim 10\%$ when adopting the higher stray light fraction, whilst the longitudinal component was amplified, typically, by several hundred gauss. All of the magnetic field values quoted throughout the paper refer to the $25\%$ stray light inversions, however, no qualitative differences ensue from the two approaches, which speaks to the robustness of the results.

The inversion was carried out over the FOV in the red box of Fig. \ref{fig:context} for the 17 minute time series. This provided, for each pixel, the vector magnetic field in spherical coordinates (i.e. the pixel-averaged magnetic field strength, $B$, the inclination of the field with respect to the LOS, $\theta$, and its azimuth in the transverse plane, $\chi$), the line-of-sight velocity, $v_{\rm LOS}$, and 5 other parameters that contain the thermodynamical information encoded in the spectral line (the source function and its gradient, the damping parameter, the Doppler width and the line-to-continuum opacity ratio).
The simplistic approximation of the Milne-Eddington atmosphere does not allow us to retrieve actual physical thermodynamical parameters, nor is it able to interpret asymmetries in the spectral line. However, it is a conservative approximation that provides a good first-order estimate of the magnetic and dynamic properties prevailing in the atmosphere \citep[e.g.][]{deltoro2016}.

In order to determine the real LOS velocity flows associated with the magnetic flux emergence, we removed the p-mode oscillations using a subsonic filter, following the space-time filtering described by \cite{title89}.
The velocities were obtained from the Level 1.2 calibrated spectra using the center-of-gravity method to determine the position of the spectral line. Then, the filter was applied over the inferred Doppler shifts. The reason to use the velocities computed with the center-of-gravity method rather than from the spectral line inversions was to avoid any spurious effect due to non-converging pixel inversions.

\section{The lower layers: photosphere}\label{photosphere}

In this section we describe the physical properties and evolution of the emergence site at the photosphere. The information presented here is inferred from the IMaX data, through spectral line inversions in the case of the magnetic properties and through the center of gravity analysis in the case of LOS velocities (see Section \ref{inversions}). The emerging magnetic field interacts with the surface granulation in complex ways, leading to morphology changes, altered plasma flows and signatures of heating.

\subsection{Vector magnetic field and LOS velocities}

\begin{figure}[!t]
\includegraphics[angle=0,scale=.37]{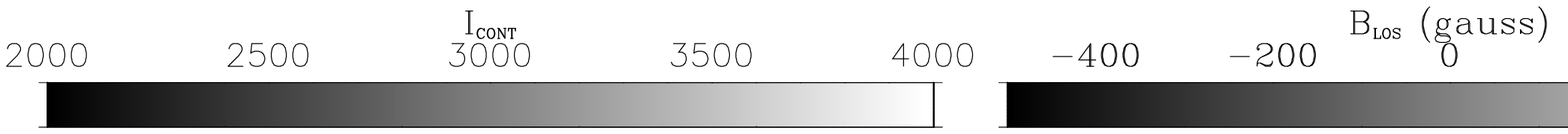}
\includegraphics[angle=0,scale=.37]{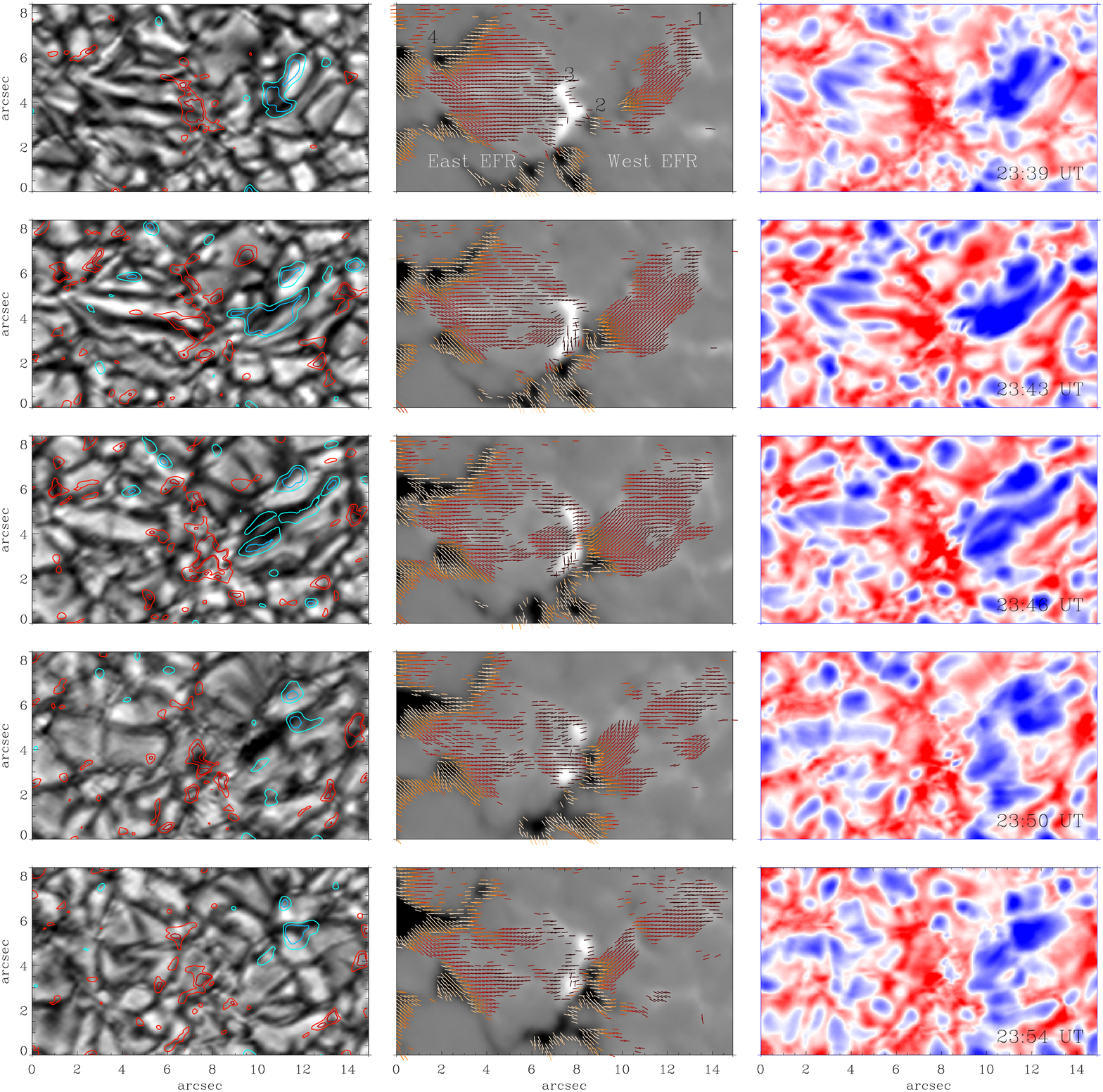}
\caption{Evolution of the two emerging flux areas inside the red box of Figure \ref{fig:context}. The left column shows the continuum intensity (at 525.043\,nm) in the background with overlaid LOS velocity contours (red and blue correspond to redshifts and blueshifts at $\pm 0.5$ and $\pm 0.8$ km/s). The middle column depicts the evolution of the magnetic field, where the background is now the LOS magnetogram and the red lines represent the transverse component of the magnetic field. In the right column, the LOS velocities after p-mode filtering are shown.\label{fig:evolution}}
\end{figure}

Figure \ref{fig:evolution} displays the evolution of the emerging flux area, sampled every six time steps. The grayscale background of the left column depicts the continuum (at 525.043\,nm) intensity images, which show the pattern of the surface granulation in the area of the emerging magnetic field. A close look reveals elongated granules ($\sim 5^{\prime\prime}$ long) during some stages of the emergence. Overplotted velocity contours at $\pm 0.8$ and $\pm 0.5$~kms$^{-1}$ represent the plasma LOS velocity, where blue and red contours correspond to material moving towards and away from the observer, respectively.
The middle column exhibits the evolution of the magnetic field. The grayscale background now represents the line-of-sight magnetic flux density, saturated at $\pm 500$~G (this quantity is obtained by computing $B\cdot cos\theta$ from the spectral line inversion results). The dashed lines show the direction of the transverse component of the magnetic field for pixels where the field strength is larger than $280$~G (this threshold was set for clarity in the figure). The color-coding of the lines changes progressively from yellow to red to black as the magnetic field inclination with respect to the surface goes from $- 40^{\circ}$ (pointing inwards) to $40^{\circ}$ (pointing outwards). Red corresponds to purely transverse magnetic field, i.e. parallel to the surface. Purely based on continuity arguments it is clear that the magnetic field comes out of the surface at the right footpoint of the West EFR (label \#1 in first row), goes back into the surface at its left footpoint (\#2), immediately threads up at the right footpoint of the East EFR (\#3) just to pierce back into the surface at the left-most footpoint (\#4). 
This column clearly shows the magnetic signature of the two emerging flux patches. The {\em confluence region} comprises labels 2 and 3 (around $x \sim 8^{\prime\prime}$ and $y \sim 4^{\prime\prime}$), where the right footpoint of the East EFR meets the left footpoint of the West EFR. Based on the connectivity of the magnetic field, this opposite polarity patch is an MDF like the ones described in \cite{bernasconi2002} and \cite{xu2010}. 

The column on the right presents the evolution of the line of sight velocity, saturated between $-0.5$ and $0.5$ km/s. Blue and red correspond to upflows and downflows, respectively. The East side emergence was already developed when IMaX started observing, while the West one was captured during its incipient stages. 
The LOS velocity shows a strong blueshift accompanying the West EFR as it rises to the surface, revealing a tight spatial correlation between horizontal magnetic fields and upflows. 
A string of downflow patches lights up the perimeter of the both EFRs during the entire sequence. The strongest and most concentrated downflows appear at the main footpoints, and especially in the confluence area. 
The surface granulation (left column) also shows the typical signatures of an emerging flux region. The presence of granules elongated in the direction of the field coincides with the presence of a strong transverse magnetic component (red lines in middle panels). As soon as this transverse component weakens, the elongated granules break down into normal quiet Sun granulation (see a more detailed analysis in Section \ref{elongated}).

These stuctures resemble emergent magnetic bubbles delimited by their footpoints that, upon ascending, perturb the granulation pattern due to the strength of their magnetic field. This is in agreement with a typical flux emergence scenario in which, as the magnetic field rises through the surface, it interacts with the convection, changing the shape of the granulation and dragging material up  into higher layers. The displaced material has to drain back down to the surface, mostly guided along the field lines to the footpoints \citep{cheung2007b, solanki2003}. 
Once the magnetic bubble has risen above the surface,  the upflows subside, the transverse fields fade and the granulation returns to its usual behaviour. In the chromospheric layers, these EFRs also resemble magnetic bubbles (see, for instance, the right panel of Fig. \ref{fig:sufi_filaments}, corresponding to SuFI's 397 nm channel).

\begin{figure}[!t]
\includegraphics[angle=0,scale=.73]{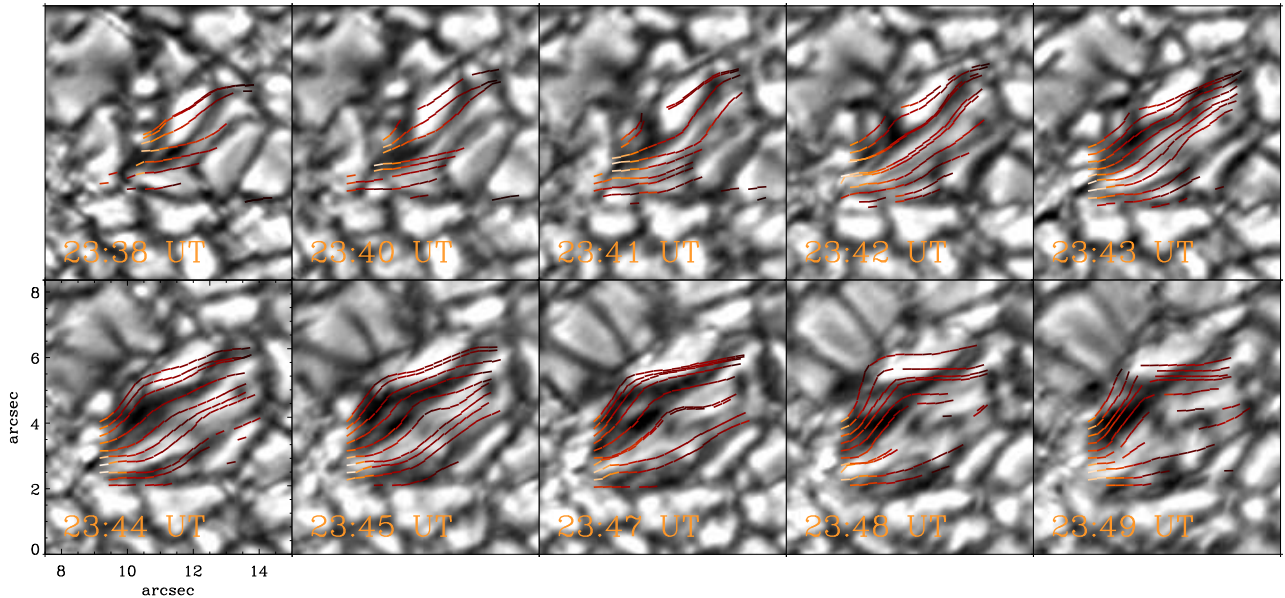}
\caption{Transverse magnetic field lines at the location of the West EFR (the spatial coordinates refer to those of Fig. \ref{fig:evolution}), which was captured from its very early stages. The background images show the continuum intensity and the red lines represent the transverse magnetic field lines connecting the footpoints of this small emerging bubble. The sequence evolves from left to right and top to bottom. \label{fig:azi_lines}}
\end{figure}

Figure \ref{fig:azi_lines} shows the evolution of the shape of the magnetic field lines at the location of the West emergence.
In this sequence (from left to right and top to bottom), the background images show the evolution of the continuum intensity. Each one of the red lines superposed on these images is constructed using a 2D version of the procedure introduced by \cite{solanki2003}: we first take a pixel in the left footpoint (\#2) and calculate the projection of the vector magnetic field onto the transverse plane (in the LOS reference frame). This is plotted over the background image with a short red dash. The direction of the vector points towards the next pixel in the sequence, which is then taken as the origin and the calculation of the transverse field is repeated. The red dashes are only drawn on the image when the magnetic field strength exceeds the threshold of 250 G. The process is iterated until the line spans the length of the emerging region. Ten ``field lines'' are represented, spanning the distance between the two footpoints (\#1 and \#2). 
It is important to emphasize that these are not magnetic field lines in the mathematical sense. They are constructed from the direction of the magnetic field in the transverse plane, which is derived from the spectral line inversions. However, they give a sense of continuity and the behavior of the magnetic field topology.

At the beginning of the sequence, only a small area is covered by magnetic field above the $250$~G threshold. In the first two panels of Fig. \ref{fig:azi_lines}, the {\em field lines} cross a few granules with no obvious alignment between them and the convective pattern. In these early moments of the emergence, the largest values of the transverse magnetic field strength (no more than $30^{\circ}$ from the transverse plane) are between 400-500\,G, which lie just below the photospheric equipartition value \citep{cheung2007b}.
As the sequence progresses, the emerging flux patch brings stronger transverse fields to the surface and widens sideways, adopting an overall rugby-ball shape. By 23:42, the transverse field strengths reach the $600-700$\,G range, increasing up to $800$\,G 2 minutes later, where they plateau and remain strong for the next 5 minutes. It is during this time that the granules adopt elongated shapes aligned with the direction of the field. Around 23:48, the transverse field strength starts to decay, as the elongated granules break up into smaller and more regular-shaped convective cells.
Based on the inferred surface field strengths, the convective motions dominate over the magnetic field at the beginning and the end of the sequence, however, in the middle, the field is well enough above the equipartition value to alter the shape of the granulation \citep{requerey2015, requerey2016}.
Towards the end of the sequence, the magnetic signal weakens at the center of the emergence (hence the lack of dashed lines in that area).  This is expected: as the field rises, it expands sideways becoming intrinsically weaker, and it moves upwards beyond the region of sensitivity of the IMaX spectral line, leading to vanishing polarization signals. 
The overall behavior though, shows field lines that fan out at the center of the emergence while remaining constrained and trapped at the footpoints, where downflows prevail. The plasma that was dragged up to the surface by the buoyant transverse fields, is funneled back down through the footpoints, guided by the field lines,  returning to the surface.

\subsection{Elongated granulation}\label{elongated}

\begin{figure}[!t]
\includegraphics[angle=0,scale=.73]{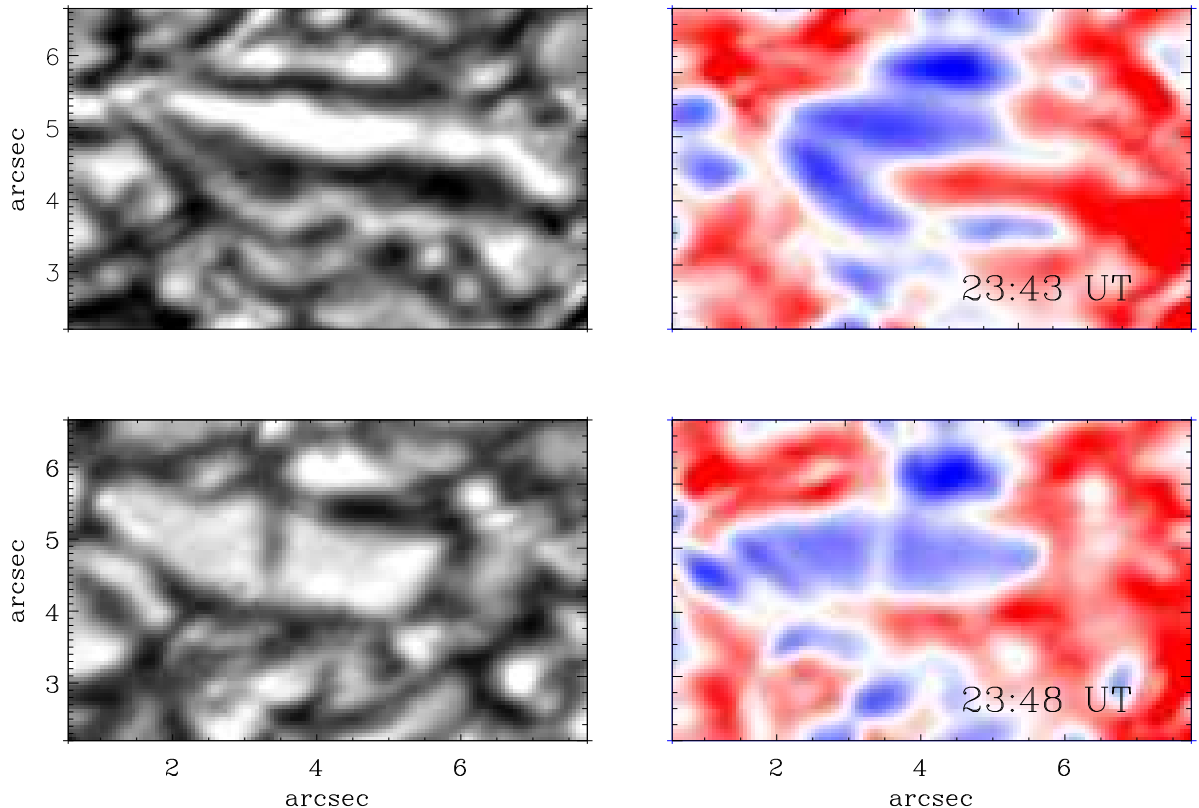}
\caption{Two snaphots in the evolution of an enlongated granule in the East EFR, seen in continuum intensity (left) and line-of-sight velocity (right) saturated at $\pm 0.8$~kms$^{-1}$. Blue indicates upflow and red downflow. The spatial coordinates refer to those of Figure \ref{fig:evolution}.  \label{fig:granule}}
\end{figure}

Elongated granules are one of the most common signatures of magnetic flux emergence in the photosphere. Figure \ref{fig:granule} shows in detail one of the elongated granules that appears during the emergence sequence: the continuum intensity on the left and, on the right, the LOS velocity saturated between $\pm 0.8$~kms$^{-1}$. The top row presents the granule when it is at its longest, coinciding with a time when strong transverse fields are present at the surface ($700 - 800$\,G). The LOS velocity for this snapshot reveals an obvious upflow pattern along the length of the bright granule while the dark elongated intergranular lane flanking it features significant downflows. The nature of the convective flow is preserved, but its morphology has changed to adapt to the presence of the magnetic field.
A few minutes later (second row), when the transverse field at the surface has weakened, the elongated structure breaks into two smaller granules, separated by a darker channel in between them. The velocity pattern changes accordingly with a neutral velocity channel appearing at the location of the rupture.

\subsection{Confluence area}\label{sec:bp}

\begin{figure}[!t]
\includegraphics[angle=0,scale=.5]{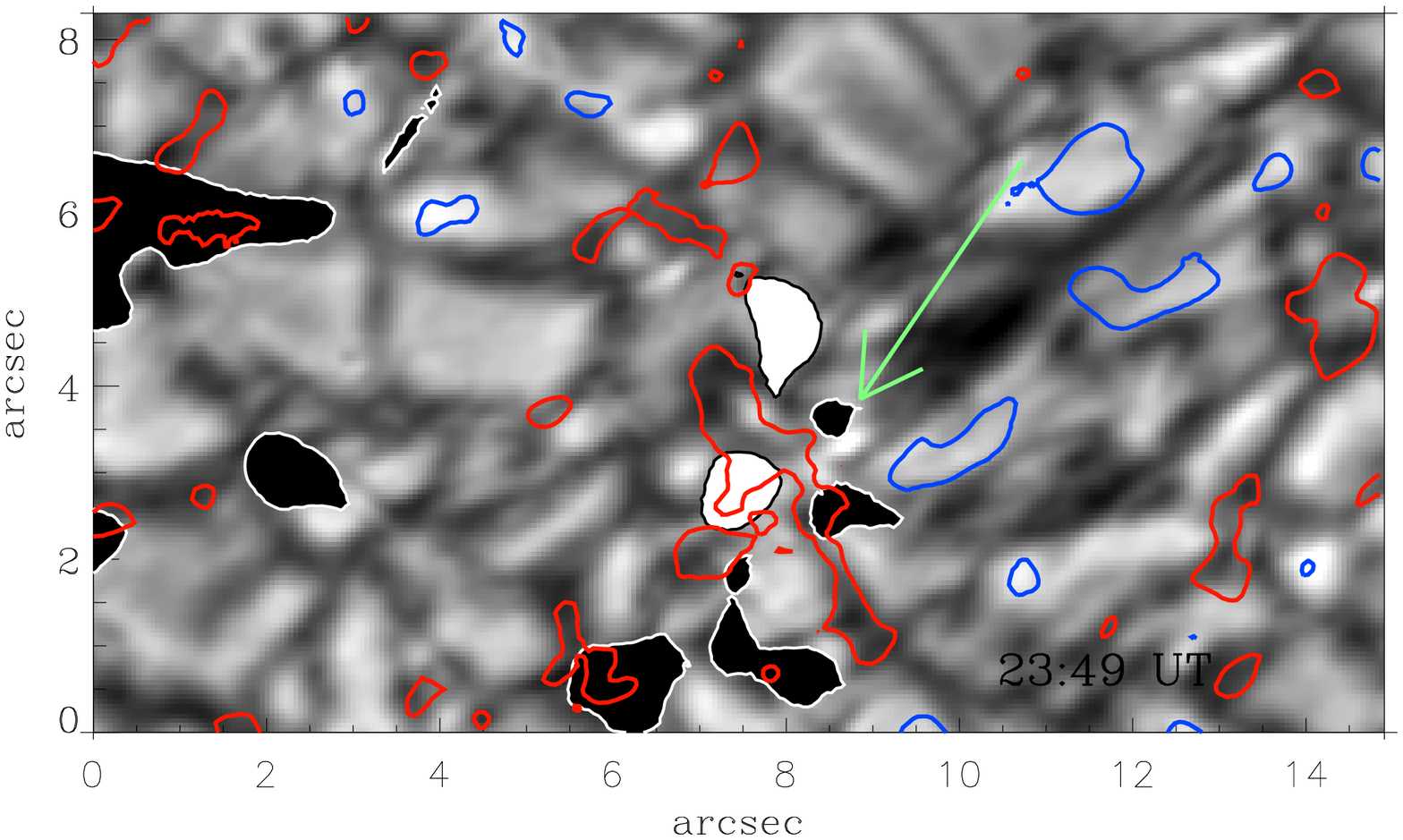}
\includegraphics[angle=0,scale=.5]{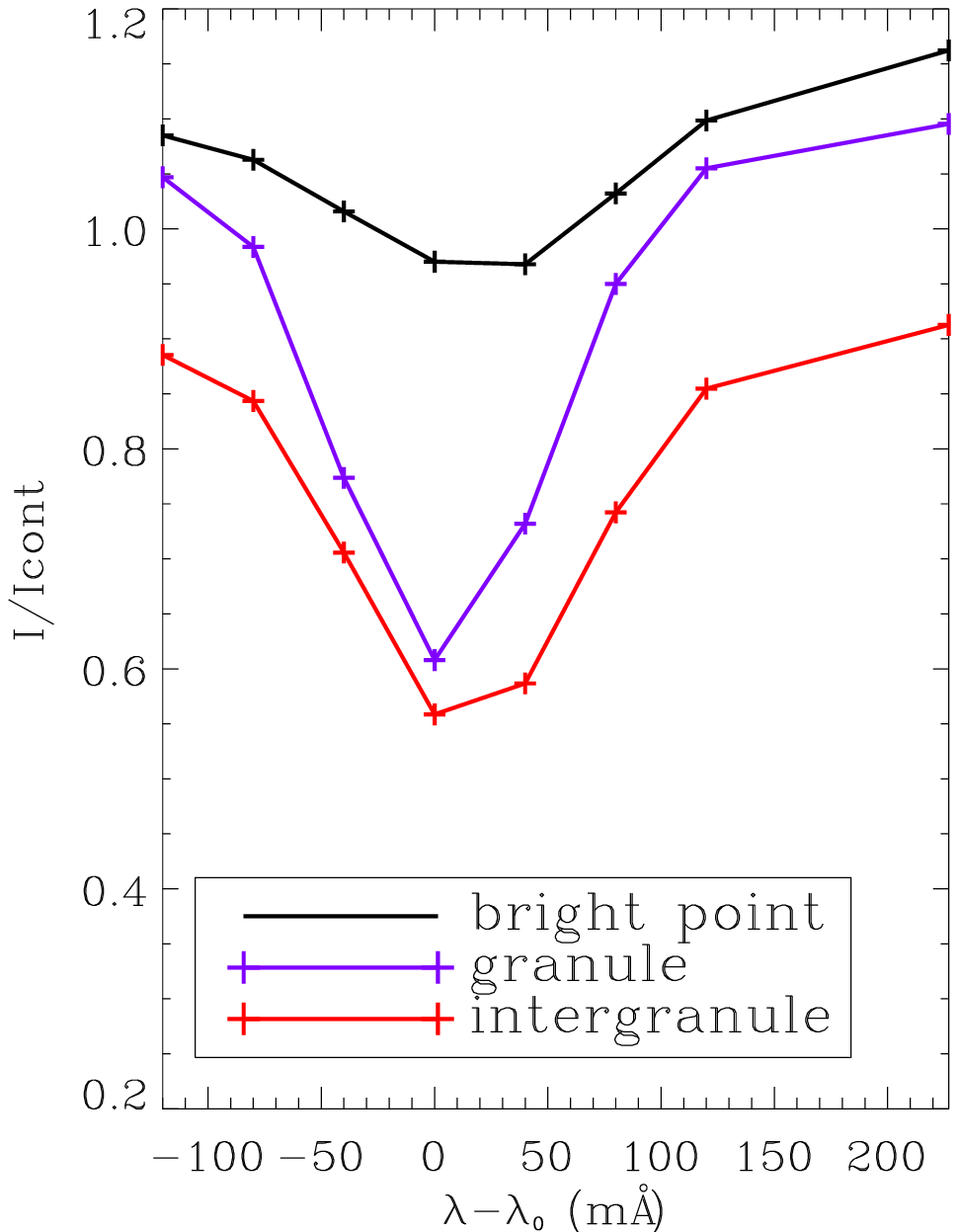}
\caption{Left: snapshot of the continuum intensity when the bright point (marked with the green arrow) in the confluence area has developed. Black and white patches show two polarities of the longitudinal magnetic field saturated at $\pm 250$~G and colored contours present the line-of-sight velocity  saturated at $\pm 0.6$~kms$^{-1}$. Blue indicates upflow and red downflow. Right: Average granular (purple) and intergranular (red) intensity profiles. An exaple of a spectral profile at the location of the bright point is shown in black. \label{fig:brightpoint}}
\end{figure}

When observing the evolution of the photospheric continuum intensity of the confluence area in between the two EFRs, a prominent bright point develops exactly at this position. It is flanked by the opposite polarity footpoints of both emerging patches, that come together as the West EFR develops and brings magnetic flux to the surface. 
The left panel of Figure \ref{fig:brightpoint} selects the instance during the lifetime of the bright point when it reaches its maximum brightness. In the background, the continuum intensity image reveals the granulation pattern as well as the bright point, whose location is indicated by the green arrow. Overplotted, the black and white patches show the longitudinal magnetic field saturated at $\pm 250$~G, whilst the blue and red contours mark the line-of-sight velocities at $\pm 0.6$~kms$^{-1}$. In this snapshot, the BP is no more than $\sim 0.5''$ wide and sits in the middle of the confluence area, where the right footpoint of the East EFR meets the left footpoint of the West EFR. It is surrounded by the strong downflows of the magnetic footpoints (red contours) and the upwelling motions of the West emerging flux region (blue contours).
The photospheric spectral profiles at the location of the bright point are very shallow in comparison to other locations in the field of view. The right panel of Figure \ref{fig:brightpoint} shows examples of a bright point (black) intensity profile, as well as the average granular (purple) and average intergranular (red) profiles. As expected, the intergranule profiles are shallow, have low continuum intensity values and are red-shifted with respect to the reference position of the spectral line. The granule profiles, on the other hand, appear blue-shifted with respect to the intergranular profiles, have a high continuum intensity and are rather deep \citep{dravins}. In the bright point, the continuum is typically $2\%-5\%$ higher than in the brightest granules, and the spectral line is wide and shallow. The spectral line seems to be slightly red-shifted, yet shows a strong asymmetry in the blue wing of the line. The higher continuum as well as the width of the spectral line indicate high temperatures and significant thermal motions, whilst the asymmetry is most likely due to a strong gradient of velocities along the line of sight. Also, the shallowness of the line indicates high temperatures in the upper layers (see section \ref{temperature_stratification} for a detailed study of the temperature statification in the FOV). The confluence of two opposite polarity footpoints and the associated chromospheric brightenings (described in Section \ref{chromosphere}) lead us to interpret the spectral profiles in the bright point as a signature of heating due to magnetic reconnection.

\begin{figure}[!t]
\includegraphics[angle=0,scale=.5]{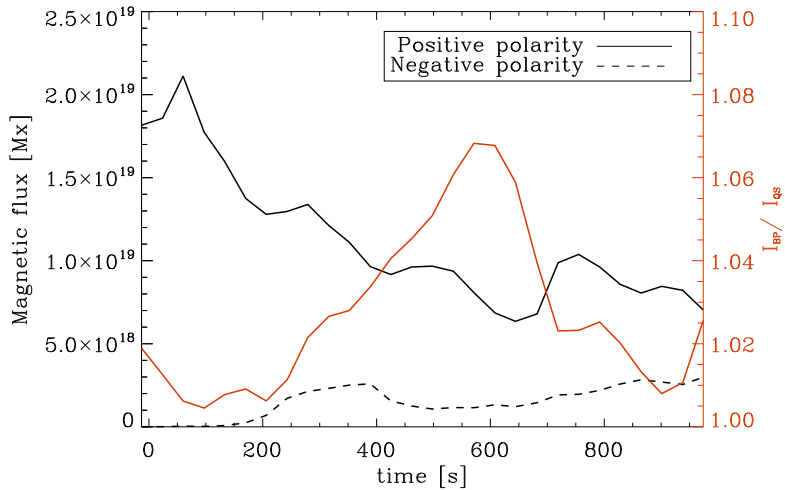}
\caption{Flux history of the confluence area (black) and light curve of bright point (red). The flux history is calculated separately for the positive polarity (solid line), which is already present at the beginning of the series, and the negative polarity (dashed line), which appears as the West EFR breaks through the surface. The threshold to calculate the fluxes is set at 300 G. The red solid line shows the light curve of the continuum intensity at the bright point, $I_{\rm BP}$, normalized to the average quiet Sun continuum intensity, $I_{\rm QS}$. \label{fig:fluxhistory}}
\end{figure}

At the beginning of the time series, the confluence area only has a patch of positive magnetic polarity, corresponding to the right footpoint (\#3) of the East EFR. As the West EFR develops, it brings flux to the surface and a negative polarity footpoint (\#2) that moves into the confluence area. Figure \ref{fig:fluxhistory} shows the flux history of the opposite polarity footpoints that come together in this small region. This is computed within a $3.3^{\prime\prime} \times 3.7^{\prime\prime}$ box with a $300$\,G threshold in order to spatially isolate the footpoints. With these constraints, the positive foopoint is always contained within the box and does not change much. Nevertheless, some positive magnetic flux does appear around $500$\,s. The negative polarity, on the other hand, enters the computation box as the West EFR brings flux to the surface. At the beginning, the only existing polarity is the positive one. But as the sequence advances, the negative flux grows as the positive flux decreases. This indicates two things: magnetic flux is being brought up to the surface by the West EFR and, as its negative footpoint moves into the confluence, the opposite polarities come together and partially cancel each other. Because the negative polarity footpoint is moving into the confluence at the same time as it cancels out with the positive polarity, its flux curve (dashed line) does not change as significantly as the positive flux curve (solid line). We can use the latter to make an estimate of the rate of flux cancellation between 50 and 650\,s, by attributing all the decay exclusively to cancellation. The positive magnetic flux loss derived from this measurement is $\sim -2.5 \cdot 10^{16}$\,Mx/s, which is 10 times larger than that found by \cite{reid2016} (but it is similar to that quoted by \cite{kuckein2012} for the flux loss rate of an entire AR). However, this is still a conservative estimate, since we are not correcting for the positive flux that enters the area during this time. The red line in the same figure shows the light curve of the bright point (calculated in a box of $1.1^{\prime\prime} \times 1.1^{\prime\prime}$ surrounding it), whose onset happens $\sim 2$ minutes after the two polarities start  cancelling out. The intensity increase stretches up to $7\,\%$ above the average quiet Sun level.

\subsection{Photospheric temperature}\label{temperature_stratification}

 \begin{figure}[!t]
\includegraphics[angle=0,scale=.5]{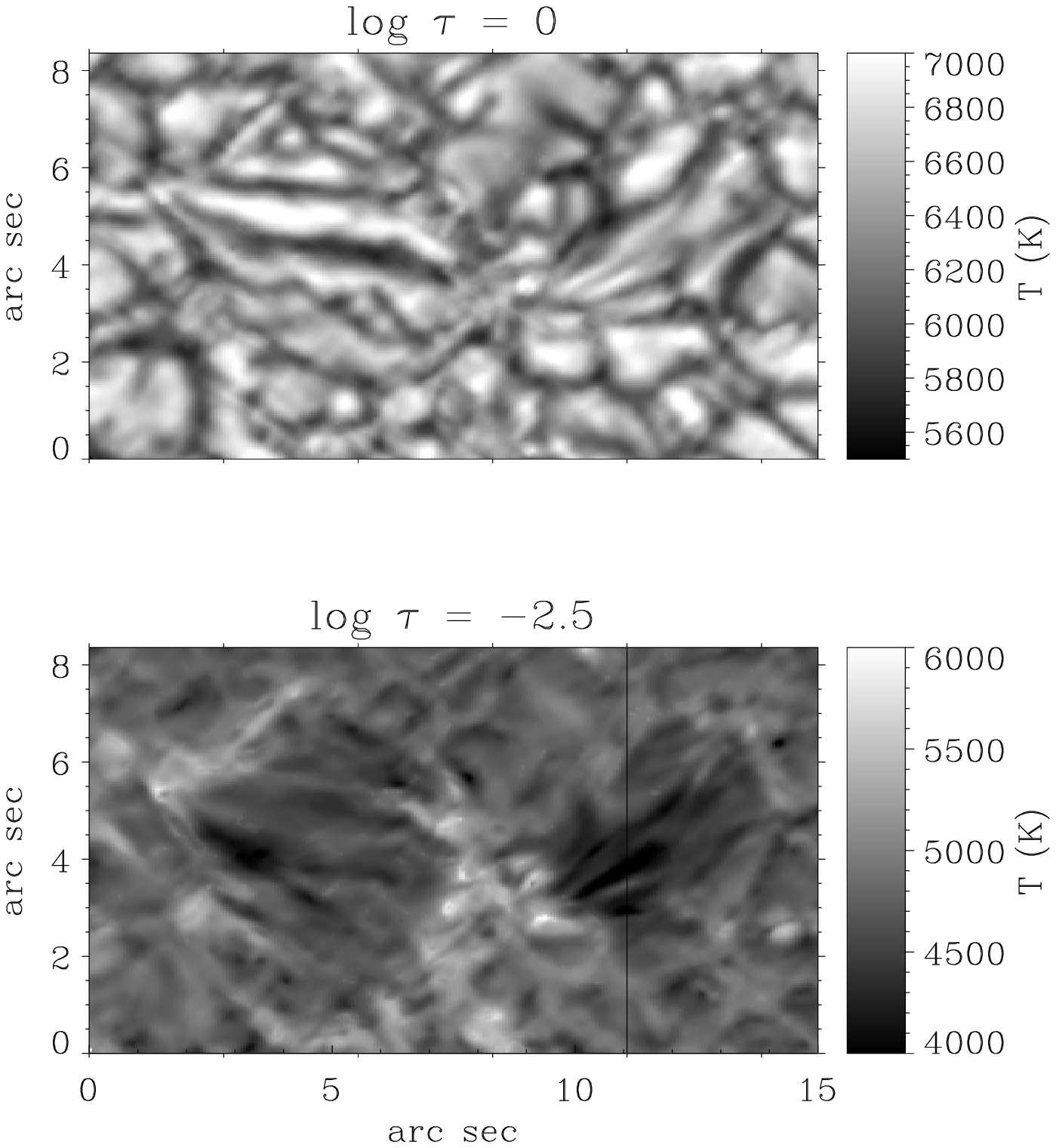}
\includegraphics[angle=0,scale=.5]{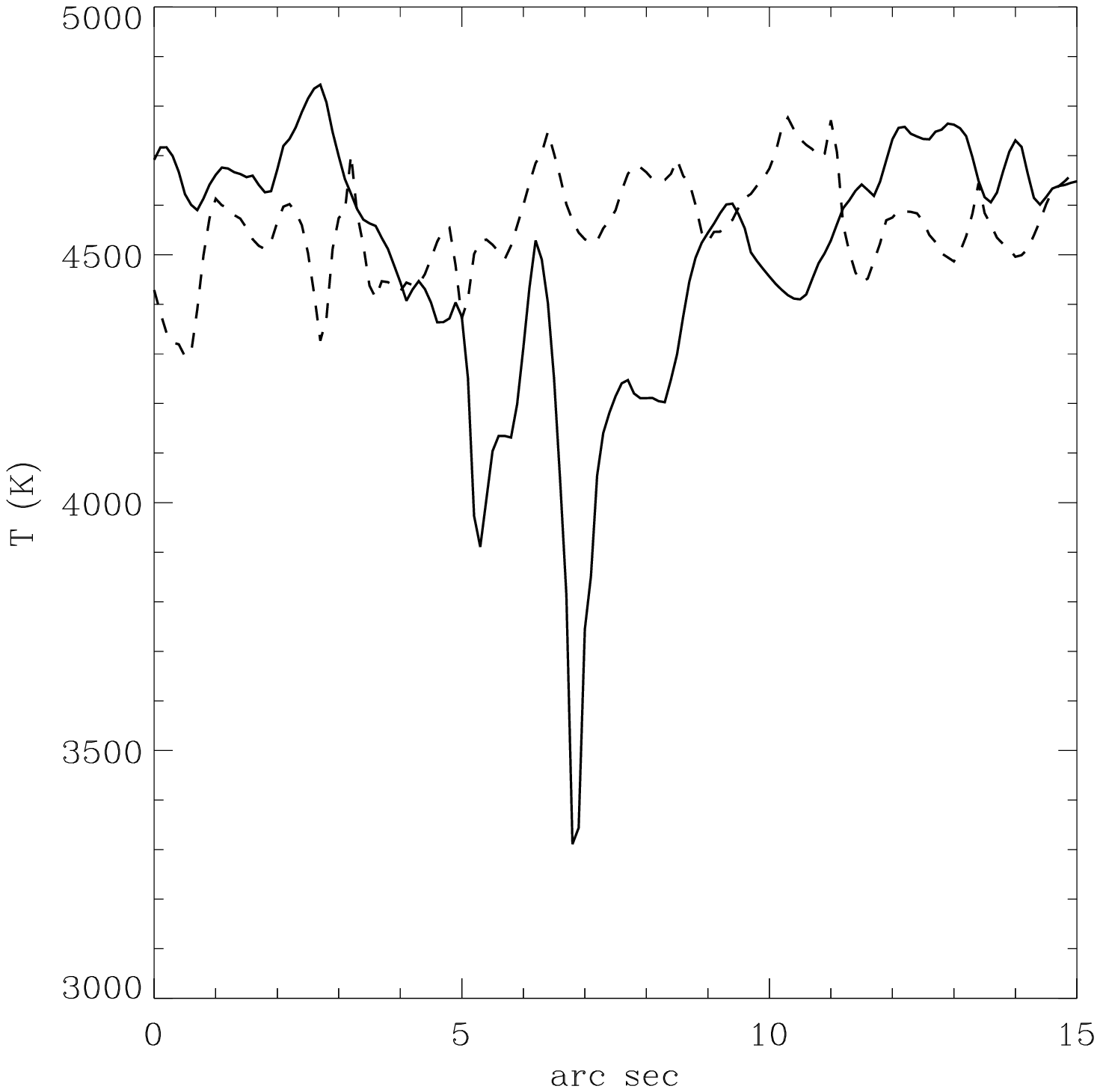}
\caption{Temperature profile at the sites of the two EFRs. Left: temperature at ${\rm log}\, \tau = 0$ and ${\rm log}\, \tau = -2.5$ for one instance in the observed sequence. Right: ${\rm log}\, \tau = -2.5$ profile along the cut in the lower left-side panel (solid) and comparison to the temperature profile at the same location but 12 minutes later (dashed). The spatial coordinates refer to those of Figure \ref{fig:evolution}. \label{fig:temperature}}
\end{figure}

One of the disadvantages of using a ME inversion code to interpret the spectral line radiation is that it does not provide certain thermodynamical variables, such as temperature or pressure. Instead, it encodes the information in the parametrization of the source function and other variables, but it does not disentangle thermodynamical information. We have used the ME approach to extract the magnetic properties in the emergence site, but in this section, we will make use of spectral line inversions in the Local Thermodynamical Equilibrium (LTE) approximation to perform an analysis of the temperature stratification.

The {\sc Sunrise} team provided LTE inversions of the Level 2 IMaX data (spatially reconstructed) carried out with the 1D version of the SPINOR code \citep[for specific details of the inversion, see][]{frutiger2000, solanki2016}. The inversion strategy applied a global stray light correction of $25\,\%$ to Stokes $I$, and allowed temperature perturbations at 3 different heights along the LOS, whilst treating the rest of the parameters (the three components of the vector magnetic field, the line-of-sight velocity and the microturbulence) as height independent.

The left side of Figure \ref{fig:temperature} shows the temperature at ${\rm log}\, \tau = 0$ (upper panel) and ${\rm log}\, \tau = -2.5$ (lower panel) for one instance in the time series. At ${\rm log}\, \tau = 0$, the temperature is highly correlated with the granulation pattern, showing hot granules surrounded by cooler integranular lanes. The elongated granulation and the bright point in the confluence area are also evident. At this layer, the confluence area experiences a temperature increase of $\sim 500$\,K with respect to its pre-bright point phase. In the lower panel, the temperature at ${\rm log}\, \tau = -2.5$ presents the expected signature of reversed granulation \citep{cheung2007a}. The two EFRs show up as cooler areas surrounded by slightly hotter ridges, especially around the footpoints. Strands of very cool material ($T < 4000$ K), roughly aligned with the direction of the magnetic field, can be seen in the magnetized areas of the two emerging patches (more so in the West one).
The solid line in the right panel of Fig. \ref{fig:temperature} shows the temperature profile along the vertical cut of the lower left panel. The dashed line corresponds to the temperature profile at the same location and height, but 12 minutes later, when the strength of the transverse magnetic field has decayed significantly. The large temperature dips captured by the solid line correspond to the strands of cool material in the West EFR. These strands appear only in the early stages of the emergence, when relatively strong magnetic field is traversing the surface layers. At the end of the sequence, the temperature profile across the EFR returns to its normal variablility expected at those heights (dashed line in right panel). 

\section{The higher layers: Chromosphere} \label{chromosphere}

\begin{figure}[!t]
\includegraphics[angle=0,scale=.35]{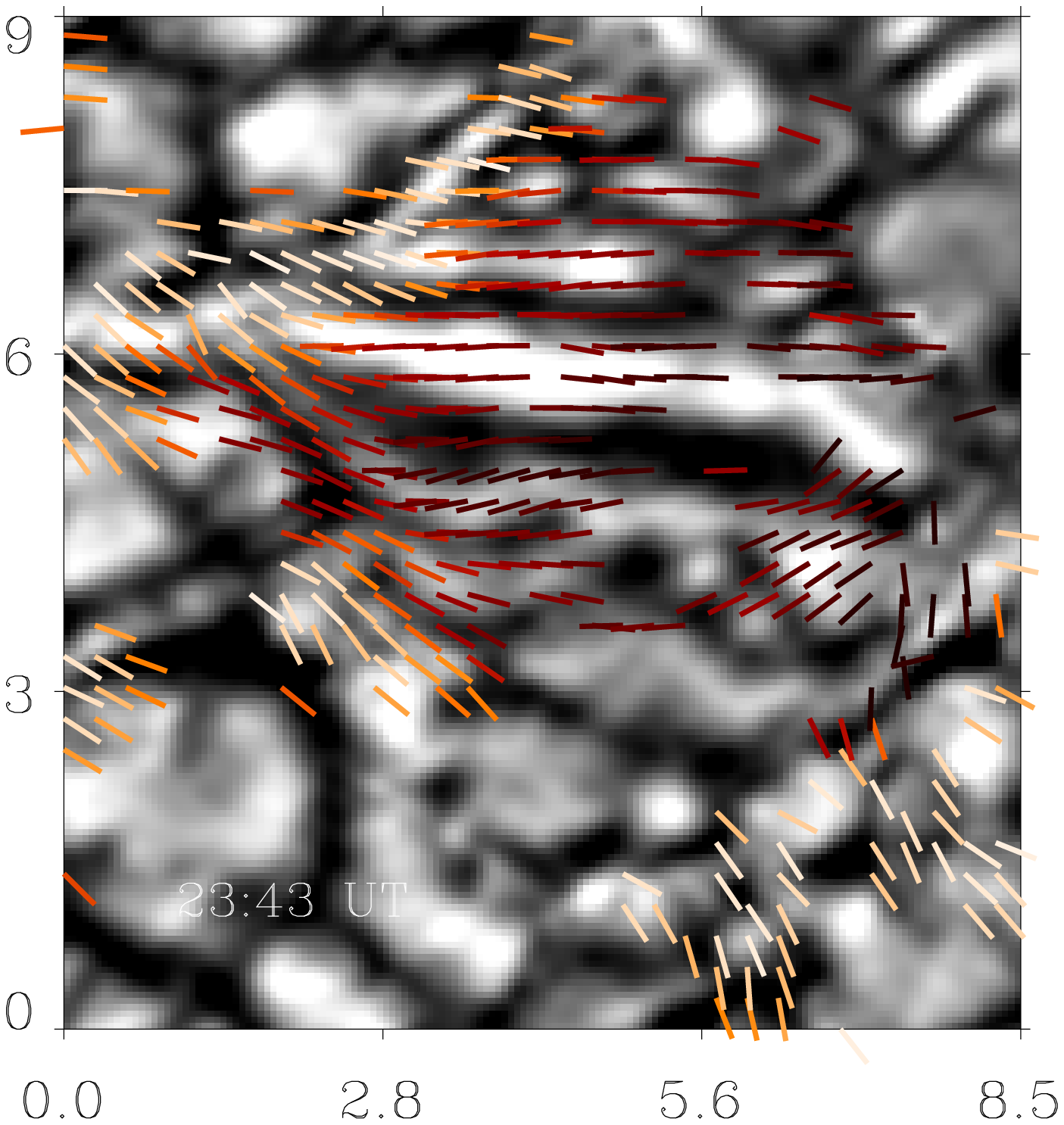}
\includegraphics[angle=0,scale=.35]{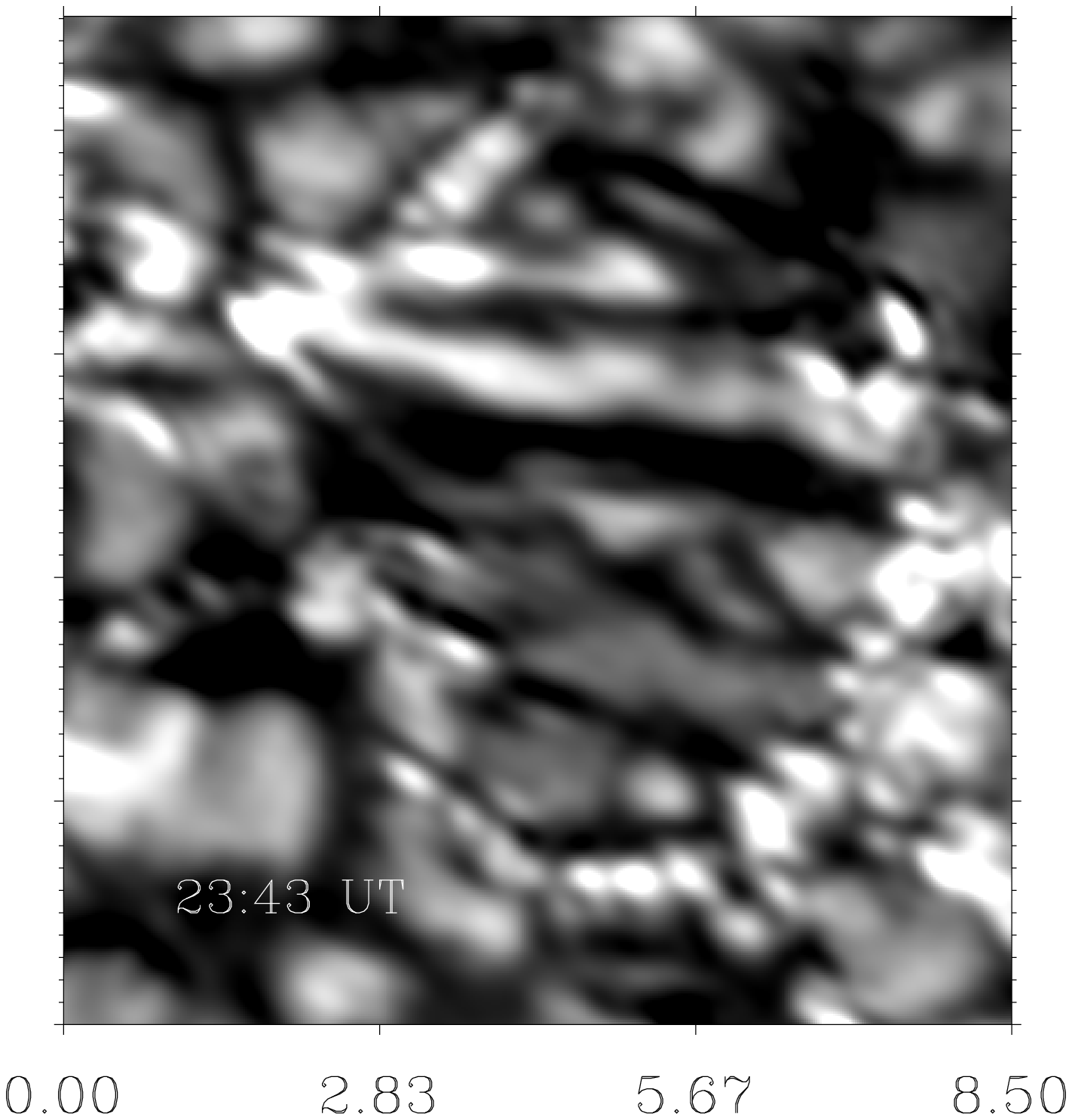}
\includegraphics[angle=0,scale=.35]{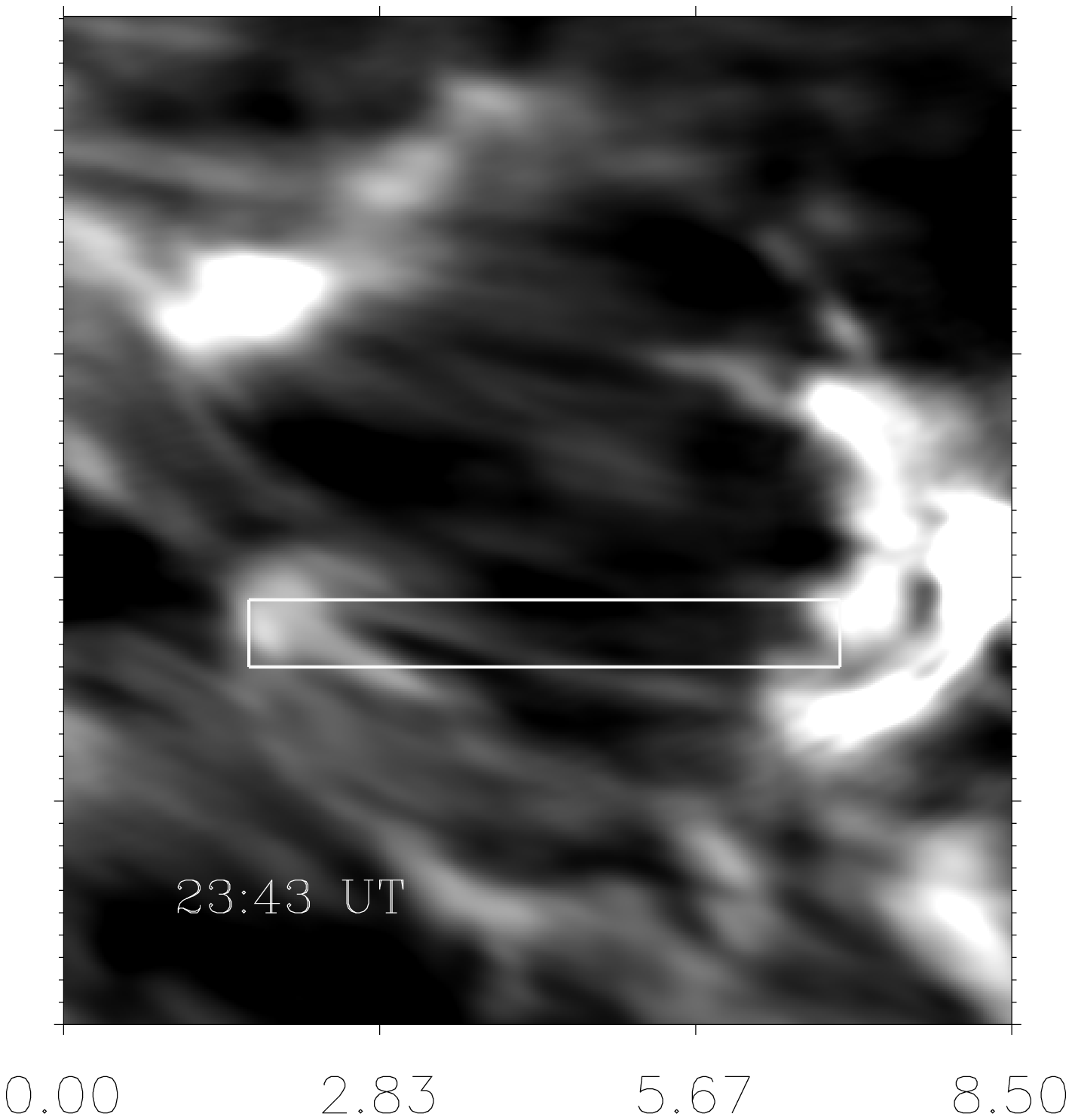}
\caption{Images of the East EFR in IMaX continuum (left), SuFI 300 nm channel (middle) and SuFI 397 nm channel (right). The IMaX image uses the same convention as the left column of Figure \ref{fig:evolution}. The UV continuum image (middle) presents a lot of similarities with the IMaX data, featuring the same elongated granule pattern, but displaying much more prominent bright points forming a chain around the flux emergence area. The chromospheric image (right) shows the confluence area as a bright region at the right edge of the figure. In this panel, the emerging flux patch appears as a dark bubble surrounded by bright points and traversed by a faint arching filament system that originates close to the footpoints. The spatial coordinates refer to those of Figure \ref{fig:evolution}, and the white rectangle is the FOV of Figure \ref{fig:fibrils}. \label{fig:sufi_filaments}}
\end{figure}

 SuFI observed the entire emergence sequence from the time that IMaX started observing, to about an hour later. Unfortunately, it only captured the East EFR and the confluence area, leaving the West EFR almost entirely outside of its field of view. SuFI recorded the event in three channels, one in the UV continuum and two bandpasses around the Ca {\rm II} H line at 396.8 nm \citep{riethmueller2013}. The latter two sampled the mid-chromosphere with significant mid-to-upper photospheric contributions \citep{danilovic2014}.  
Figure \ref{fig:sufi_filaments} shows images of the East EFR in different wavelength channels. At photospheric layers, the IMaX continuum (left) and the UV continuum (middle) show the expected signatures of flux emergence: elongated granules flanked by dark lanes aligned in the direction of the magnetic field. The main difference between these two images is that the UV continuum shows a string of brighter points surrounding the slightly darker EFR. At chromospheric layers (right panel) the emerging flux appears clearly delineated by bright points surrounding a darker cell interior. The dark bubble, in turn, is traversed by a faint filament system that spans from footpoint to footpoint, along the direction of the magnetic field inferred from IMaX data (which is shown by the red dashes in the left panel, using the same convention as in Figure \ref{fig:evolution}). The confluence area is the brightest region located close to the right border of the image (this is also the edge of the SuFI FOV). 

\subsection{SuFI Ca{\rm II} 397 nm}

\begin{figure}[!t]
\includegraphics[angle=0,scale=.45]{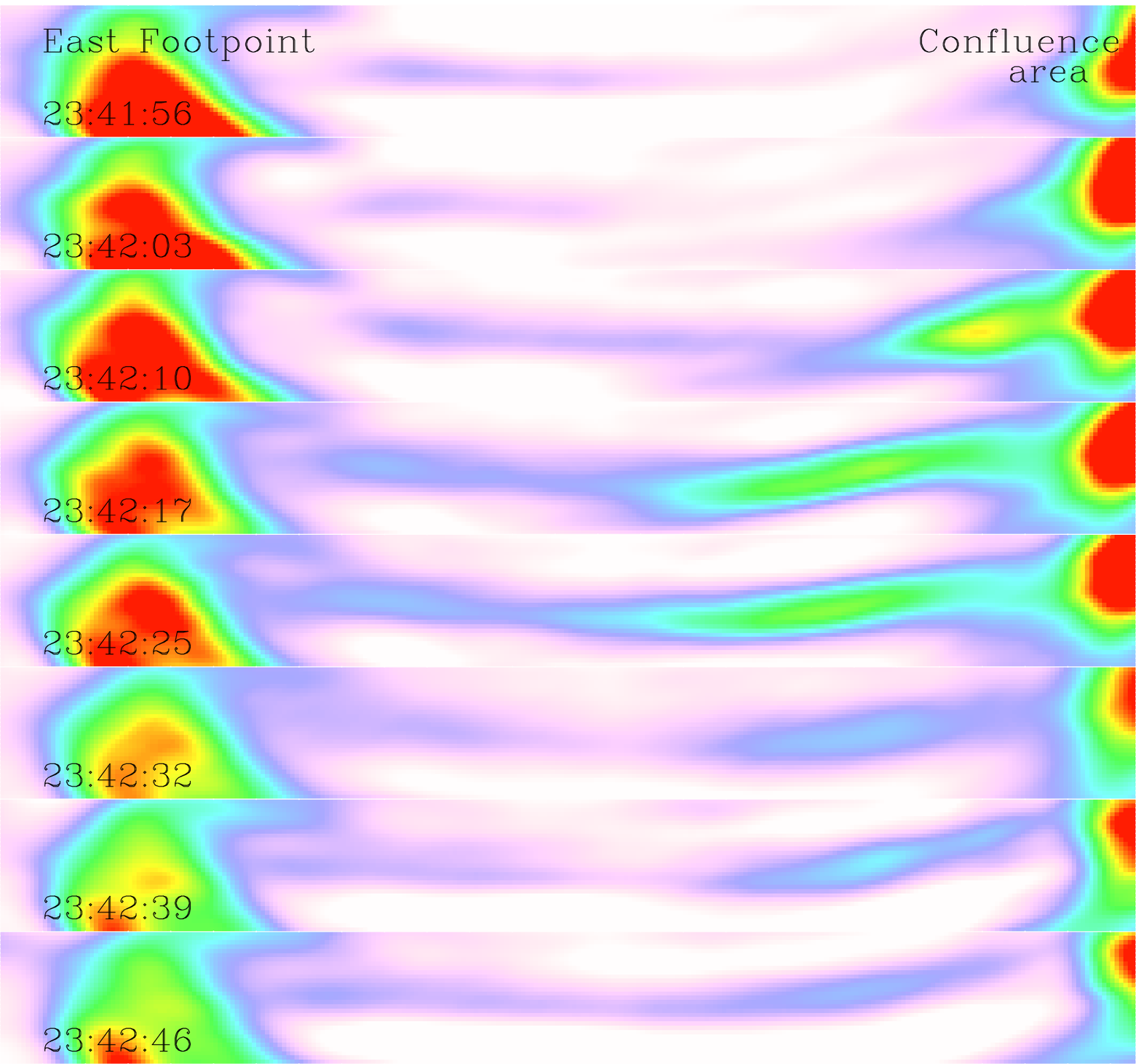}
\includegraphics[angle=0,scale=.45]{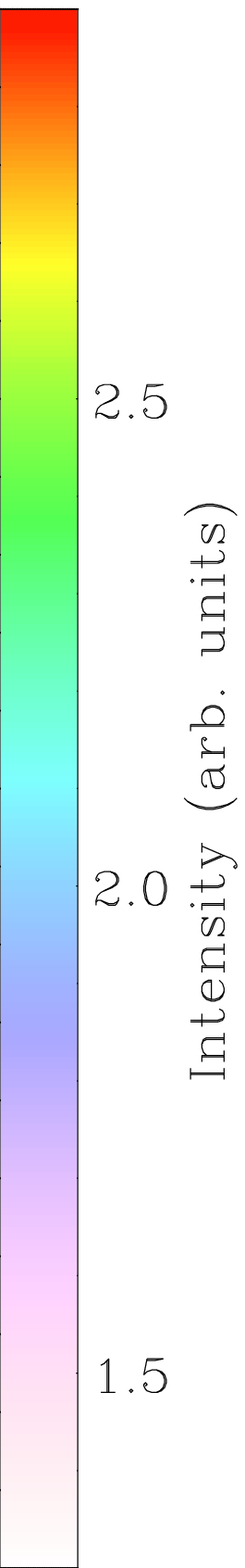}
\caption{Time sequence of SuFI 397 brightness image inside the white rectangle in the right panel of Figure \ref{fig:sufi_filaments} (although not co-temporal with it). The left side of each panel corresponds to the East-most footpoint, whilst the right side ends in the confluence area. A finger-like brightening protrudes from the confluence area and extends towards the East footpoint, becoming longer with time and then fading away again. \label{fig:fibrils}}
\end{figure}

The SuFI images at 397 nm sample the emerging magnetic flux at mid Chromospheric layers. At the beginning of the observing sequence, the East EFR appears as a dark bubble surrounded by bright edges. But as time goes by, an arch filament system develops above the region, with bright filaments that span between the two footpoints. The filaments start to appear as the bright point in the confluence area develops. In fact, some of the filaments seem to protrude from this region and elongate towards the left-most footpoint (footpoint \#4 in Fig. \ref{fig:evolution}), as depicted in Figure \ref{fig:fibrils}. There are many instances of this behavior, i.e., brightenings that start at the confluence footpoint and apparently travel towards the left-most footpoint of the East EFR (\#4). Because the FOV of SuFI ends at the confluence, we cannot know if the same happens over the West EFR, i.e. chromospheric fibrils developing from the confluence  towards the right-most footpoint (\#1). The absence of spectroscopic measurements renders impossible the attribution of this apparent motion to outflows from the confluence point, however, it is in any case indicative of energy release, whether in the form of motion or heat (or both).


\subsection{AIA UV continuum at 160  and 170 nm}

\begin{figure}[!t]
\includegraphics[angle=0,scale=.5]{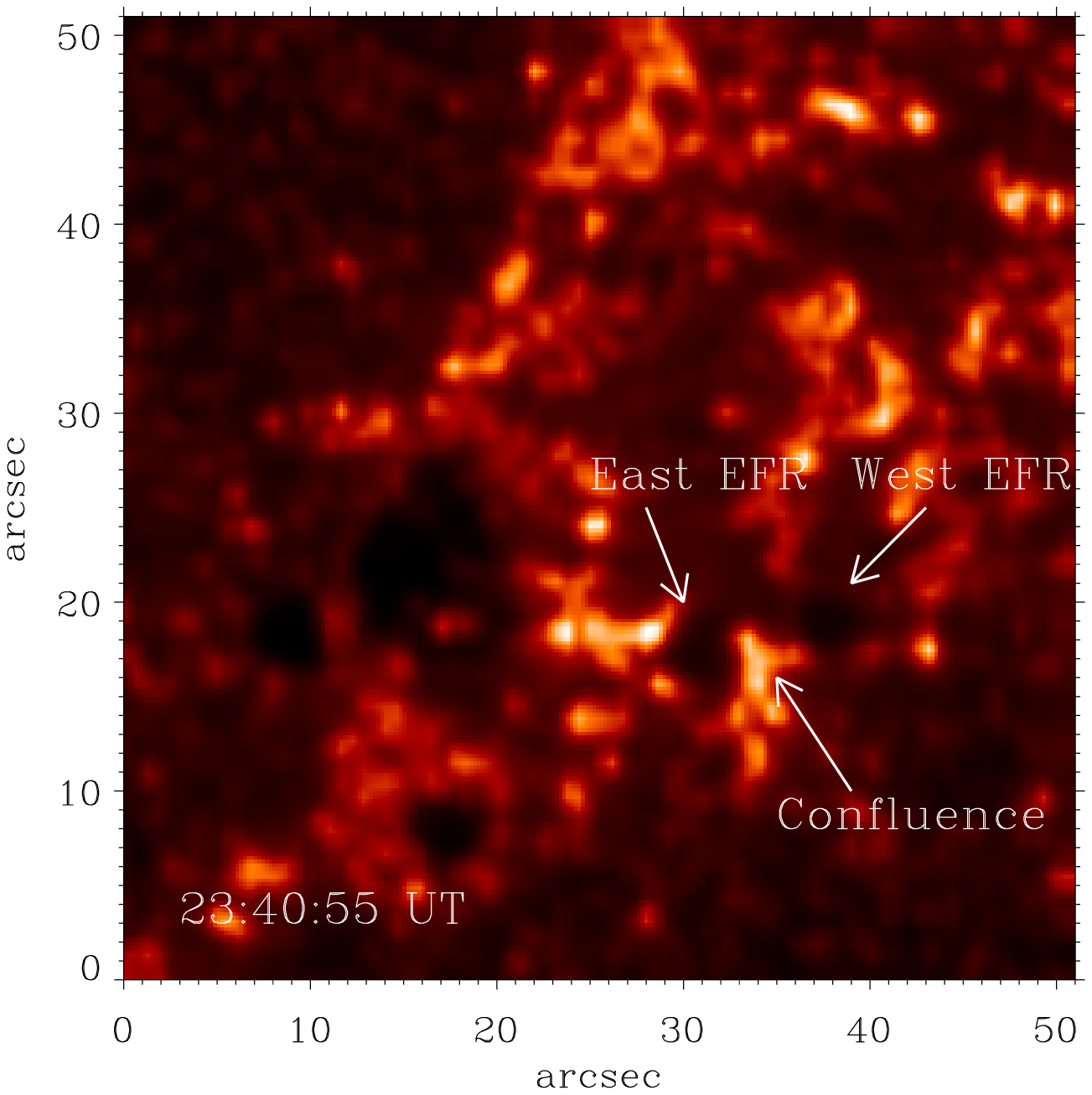}
\includegraphics[angle=0,scale=.47]{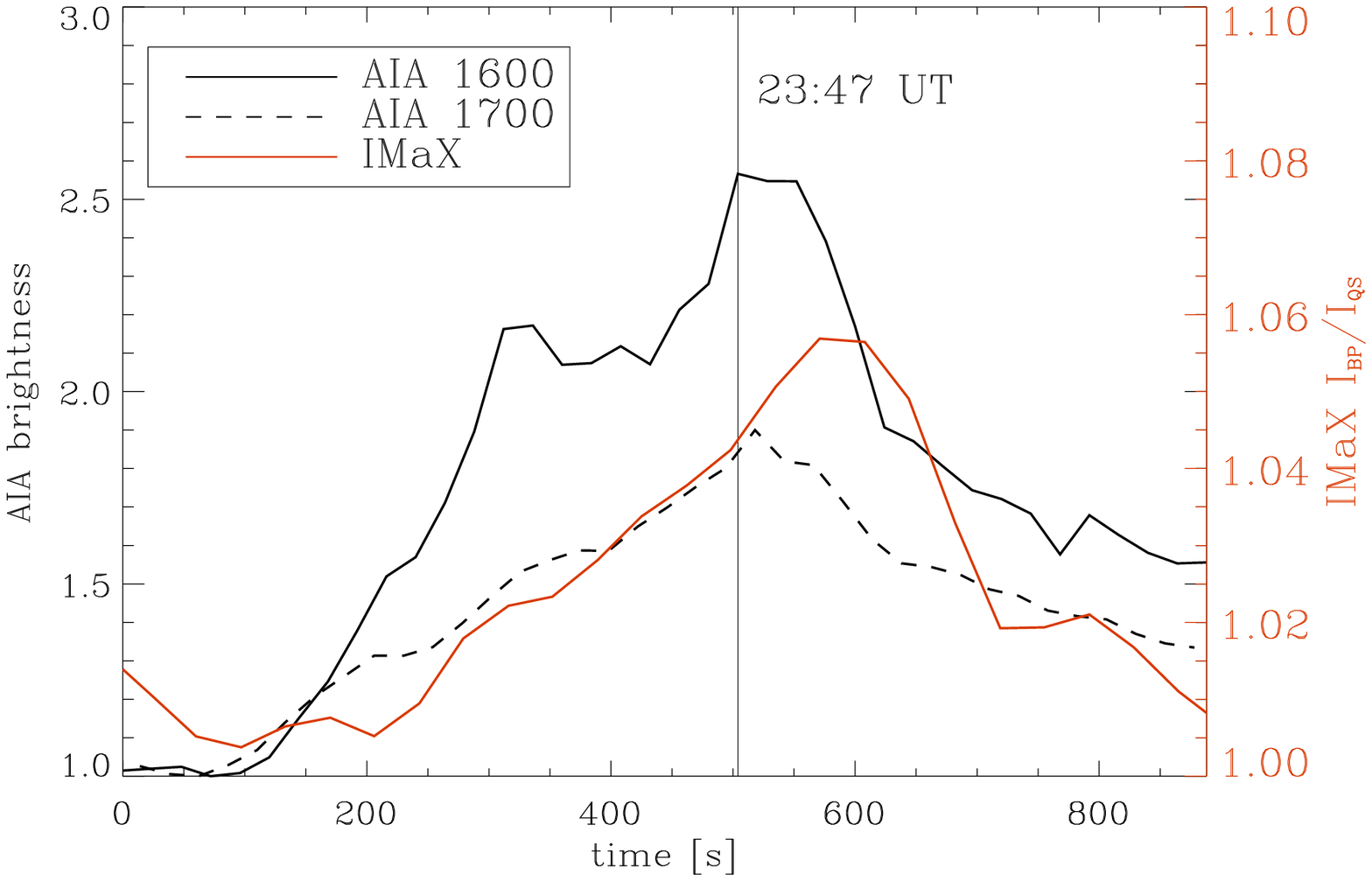}
\caption{AIA 160 and 170 nm observations. The left panel shows an image of the full IMaX FOV (compare to Fig. \ref{fig:context}) in the 160 nm channel of AIA, two minutes after the beginning of the {\sc Sunrise} observations. The right panel shows light curves of the two UV channels in a $4.8^{\prime\prime} \times 4.8^{\prime\prime}$ box around the confluence point. The light curve for the IMaX bright point is shown in red for comparison. \label{fig:aia}}
\end{figure}

The AIA \citep{aia_paper} instrument on board SDO also registered the emergence events. Despite the lower spatial resolution, the UV continuum channels at 160 and 170 nm show the brightenings delimiting the small emerging flux regions. This allows us to unequivocally identify them in the UV images. The left panel of Fig. \ref{fig:aia} shows an AIA 160 nm image, cut out to approximately reproduce the IMaX FOV, two minutes into the {\sc Sunrise} observations. Both emerging regions appear as dark bubbles surrounded by bright dots. The confluence area between the opposite polarity footpoints, also shows excess brightness. Throughout the 17 minute observation, the UV emission in the confluence area increases rapidly in both channels, developing a bright point that completely dominates the emission in the entire FOV. The right panel in Fig. \ref{fig:aia} shows the light curves in the 160 nm and 170 nm channels. The light curves were computed by integrating the brightness in a small $4.8^{\prime\prime} \times 4.8^{\prime\prime}$ box around the confluence area and were both normalized to its quiet Sun value. The brightness starts to increase approximately 2 minutes after the beginning of the IMaX observations and reaches its maximum 8 minutes later, increasing by a factor of 1.8 in 170 nm and by 2.6 in 160 nm.

As described in Section \ref{sec:bp}, a bright point in the IMaX continuum image also appears in the confluence area.
The IMaX continuum light curve for this event (red solid line) has its onset about 100 s after the onset of the AIA event, and reaches its maximum intensity around the time that the AIA brightness starts to decay. Due to the lower cadence of the IMaX data, the time lag of the visible with respect to the UV brightening is subject to a large uncertainty, but the visible bright point lags the UV one by at least half a minute.

\section{Discussion} \label{discussion}
 
The widely accepted picture of the emergence of Active Regions invokes the rise of an $\Omega$-loop from a toroidal magnetic field that sits near the base of the convection zone, triggered by the buoyancy instabilities. When the field reaches subphotospheric layers it interacts with the convective flows, suffering a distortion that depends on the original strength and twist of the flux tube \citep{cheung2007b}. This leads to a systematic undulation along the length tube (with typical wavelengths of 1 -2 Mm), where upflows aid the rise of the crests whilst downflows supress the rise of the troughs \citep{pariat2004, cheung2010}.
During the first day of its 2013 flight, {\sc Sunrise} witnessed the rise of magnetic flux at the emergence site of an AR. The evolution of very small-scale processes was recorded in detail by the two onboard instruments: IMaX and SuFI. This enabled the retrieval of the full vector magnetic field in the photosphere and the chromospheric response to the flux emergence. 

 {\sc Sunrise} captured the evolution of 2 small EFRs during the 17m observation (see Fig. \ref{fig:evolution}). Each one comprises a pair of opposite polarity footpoints connected by a horizontal magnetic field which is roughly aligned with the axis of the EFR and the polarity of the large-scale AR. The positive footpoint of the East EFR (\#3 in Fig. \ref{fig:evolution}) is contiguous to the negative footpoint of the West EFR, in an MDF configuration. Continuity arguments suggest a field topology that emerges from under the surface at \#1, sinks at \#2, threads up again through \#3 and returns back through the surface at \#4, in an undulating fashion. The velocities are such that all foopoints experience downflows whilst the horizontal field patches present upflows. 
 The East EFR is already developed by the time the observation starts, while the West EFR is captured from its initial stages. At the beginning of its emergence process, the transverse field at the surface is weak ($<500$\,G) and connects the footpoints without distorting the granulation. As stronger field ($>700$\,G) rises up to the surface, the granulation adopts an elongated pattern aligned with the direction of the horizontal field. This lasts for $\sim 5$\,m, roughly a granule lifetime, after which the strong upflows cease, the transverse magnetic field weakens and the convective motions take charge again, creating slight dips in the field at the intergranular lanes. This results in the breakage of the elongated convection cells, returning to a granulation pattern that resembles that of the quiet Sun.  The entire process happens in just over 10 minutes, which is similar to the normal granular time scale. 
 
 Whether there is connectivity of the magnetic field below the surface between footpoints \#2 and \#3 is, of course, debatable. There is no evidence in the data aside from proximity and magnetic field continuity arguments. However, this small region where the two footpoints meet (referred to throughout the paper as the {\em confluence area}), is the stage for some interesting happenings that take place in the following order:

  \begin{tikzpicture}[snake=zigzag, line before snake = 5mm, line after snake = 5mm]
    \draw (0,0) -- (14,0);

    \foreach \x in {0,2,4,6.6,8,10,11, 14}
      \draw (\x cm,3pt) -- (\x cm,-3pt);

  	\draw (0,0) node[below=3pt] { 0s } node[above=3pt] {   };
    	\draw (2,0) node[below=3pt] {Flux } node[above=3pt] {starts};
        \draw (2,0) node[below=14pt] {cancellation} node[above=14pt] {AIA BP};
                \draw (2,0) node[below=25pt] {starts} node[above=25pt] {};
    \draw (4,0) node[below=3pt] {} node[above=14pt] {IMaX BP};
        \draw (4,0) node[below=3pt] {200 s} node[above=3pt] {starts};
    \draw (6.6,0) node[below=3pt] {} node[above=3pt] {system};
        \draw (6.6,0) node[below=14pt] { } node[above=14pt] {filament};
                \draw (6.6,0) node[below=25pt] { } node[above=25pt] {Arch};
                \draw(8,0)node[below=3pt] {400s}; 
        \draw (10,0) node[below	=14pt] { } node[above=14pt] {AIA BP};
    \draw (10,0) node[below=3pt] {  } node[above=3pt] {max};
    \draw (11,0) node[below=3pt] {IMaX BP} node[above=3pt] {};
        \draw (11,0) node[below=14pt] {max} node[above=14pt] {};
    \draw (14,0) node[below=25pt] {} node[above=25pt] { Flux };
        \draw (14,0) node[below=24pt] {} node[above=14pt] { cancellation};
                \draw (14,0) node[below=3pt] {700 s} node[above=3pt] { ends};
  \end{tikzpicture}

At the beginning of the observation the East EFR is fully developed. Its positive footpoint (\#3) sits alone in the confluence region and strong downflows prevail in the area. The West EFR is surfacing, but still far from the confluence.
About $100$\,s later, the West EFR brings more magnetic flux up to the surface accompanied by strong upflows. At this time, its negative polarity footpoint (\#2) develops and moves into the confluence area, and magnetic flux starts to cancel out. Almost simultaneously, a bright point (BP) develops in the 160 and 170 nm channels of AIA.
Around $200$\,s a bright point also starts to develop in the visible continuum. Later, at $330$\,s, an arch-filament system over the East EFR becomes visible in SuFI's 396.8 nm channels, and apparent outflows that start at the confluence area seem to travel to the East-most footpoint (\#4).
The AIA bright point reaches its maximum brightness at around $500$\,s, followed by its analogous event in the IMaX continuum a few seconds later. At $700$\,s the flux cancellation at the photosphere tapers off, the transverse magnetic field at the surface has weakened notably in both EFRs and the granulation has almost returned to its quiet Sun pattern. The chromospheric images, however, still show a bundle of bright strands overlying the East EFR.

The magnetic flux cancellation and the ensuing photospheric and chromospheric brightenings strongly suggest that magnetic reconnection is taking place at the confluence point. The flux removal process is expected to show transverse field signatures at the neutral line between the cancelling polarities, however, the spatial resolution of the IMaX data might still not be enough to see this \citep[see][for an analysis of expected polarization signatures at flux cancellation sites]{kubo2014}. The downflows in the photosphere and the apparent outflows from the confluence into the arch filament system above the East EFR suggest bidirectional flows and/or energy transfer originating somewhere in the upper photosphere, rather than below the surface. This is supported by the asymmetric photospheric spectral profiles (see Fig. \ref{fig:brightpoint}) and the fact that the bright point in the visible continuum appears later than the bright point in the UV continuum and SuFI's 396.8 nm band. Unfortunately, the lack of spectroscopic measurements in Ca {\sc ii} H prevent us from attributing the apparent chromospheric outflows to actual plasma motions.

\cite{vissers2015} argue that the AIA 160 and 170 nm channels are great indicators to identify EBs, since they show up as brightenings of a point-like nature (rather than filamentary strands). However, \cite{rutten2016} states that they should not be visible in the optical continuum. Rather, that aside from the Balmer lines, other spectral indicators of EBs are brightenings in strong Ca {\sc ii} H \& K, other UV lines and UV continuum. 
All our observations point towards an EB happening at the confluence: 1. emergence of new flux in a developing AR; 2.  there is an MDF (bipole of the opposite polarity to that of the AR; the MDF harbors a magnetic U-loop under the surface) at the confluence; 3. magnetic flux cancellation takes places in the MDF; 4. strong brightnings ensue in the UV continuum and Ca {\sc ii} H; 5. there are downflows at the photosphere and apparent outflows in the chromosphere, suggesting bi-directional flows; 6. the photospheric temperature increases by $\sim 500$\,K at the location of the bright point. All of these indicate that the reconnection starts somewhere in the upper photosphere and travels down towards the surface, where the flux cancellation is visible in the magnetograms.
The absence of H$_{\alpha}$ spectra in our dataset prevents us from establishing whether the confluence point is host and witness to an EB; however, the magnetic flux cancellation and the photospheric and chromospheric responses are, strong indicators of reconnection and heating.

EBs are estimated to release between $10^{23}$ and $10^{26}$ ergs of radiative energy per event \citep{reid2016} and have typical durations of 10 minutes \citep{georgoulis2002}. During the emergence of ARs, MDFs are a common occurence. They take place every time the field lines that span the distance between the main footpoints of the AR dip below the surface. Estimates from several authors \citep{pariat2004, centeno2012} suggest that this happens every 5 to 10$^{\prime\prime}$ ($\sim 3 - 8 $\,Mm), and several strings of MDFs can exist in parallel, along the main direction of the AR. As the distance between the main footpoints of the AR becomes larger, more MDFs will form whenever new flux emerges to the surface, and each one of them is a likely scenario for an EB to happen (\cite{pariat2004} associated $\sim70\%$ of their EBs to dipped U-loops). In this way, magnetic field reconnection naturally takes place in the resistive emergence scenario of developing ARs, and could account for a lot of the heating seen in the higher layers.

\section{Conclusions}

To our knowledge, the {\sc Sunrise II} data presented in this paper have provided the most detailed observation of the small-scale emergence of magnetic flux in developing ARs to date, unveiling the topology of the full vector magnetic field at the photosphere and the response of the low chromospheric layers. Signatures of reconnection (magnetic flux cancellation and ensuing photospheric and chromospheric brightenings, temperature enhancements and possible bi-directional flows) take place around the U-loops where the emerging magnetic field remains trapped below the surface. This is a common occurrence during the formation phases of ARs that could account for a lot of the heating in and around the arch filament systems seen at higher layers.

\begin{acknowledgements}
The National Center for Atmospheric Research is sponsored by the National Science Foundation.The German contribution to \textsc{Sunrise} and its reflight was funded by the Max Planck Foundation, the Strategic Innovations Fund of the President of the Max Planck Society (MPG), DLR, and private donations by supporting members of the Max Planck Society, which is gratefully acknowledged. The Spanish contribution was funded by the Ministerio de Econom\'ia y Competitividad under Projects ESP2013-47349-C6 and ESP2014-56169-C6, partially using European FEDER funds. The HAO contribution was partly funded through NASA grant number NNX13AE95G. This work was partly supported by the BK21 plus program through the National Research Foundation (NRF) funded by the Ministry of Education of Korea. The HMI and AIA data used in this work are courtesy of NASA/SDO and the HMI and AIA science teams. The National Solar Observatory (NSO) is operated by the Association of Universities for Research in Astronomy (AURA) Inc. under a cooperative agreement with the National Science Foundation. 

\end{acknowledgements}

\clearpage

\end{document}